\documentstyle[epsfig,12pt]{article}
\begin{document}



\begin{center}
{\large{\bf Generalized Fock spaces, new forms of quantum }}

\vspace{.2cm}

{\large{\bf statistics and their algebras}}

\vspace{1cm}

{\bf A.K.Mishra and G.Rajasekaran} 

\medskip

{\it Institute of Mathematical Sciences, } \\
{\it Taramani, Madras-600 113} \\
{\it INDIA.} \\
{\it e-mail:~mishra@imsc.ernet.in} \ , \ {\it
graj@imsc.ernet.in}

\end{center}

\vspace{.3cm}

\baselineskip=18pt
{\centerline{\bf Abstract}}

We formulate a theory of generalized Fock spaces which underlies the
different forms of quantum statistics such as ``infinite'', Bose-Einstein
and Fermi-Dirac statistics. Single-indexed systems as well as
multi-indexed systems that cannot be mapped into single-indexed systems
are studied. Our theory is based on a three-tiered structure consisting
of Fock space, statistics and algebra. This general formalism not only 
unifies the
various forms of statistics and algebras, but also allows us to
construct many new forms of quantum statistics as well as many algebras
of creation and destruction operators. Some of these are : new algebras
for infinite statistics, q-statistics and its many avatars, a consistent
algebra for fractional statistics, null statistics or statistics of
frozen order, ``doubly-infinite'' statistics, many representations of
orthostatistics, Hubbard statistics and its variations.

\vspace{.3cm}

\noindent {\bf Keywords :} \ Fock spaces ; quantum statistics ;
$q$-deformations ; quantum groups ; Hubbard model ; orthostatistics.

\vspace{.3cm}

\noindent {\bf PACS Nos:} \ 03.70 ; 05.30 ; 02.20 ; 71.27

\vspace{.3cm}


\newpage


\baselineskip=24pt
\setcounter{page}{1}
\noindent{\bf {1. Introduction}}

Recently, much effort has been devoted to q-deformations of
oscillators.  Both single oscillators [1-12] as well as systems of oscillators
[13-33] have been studied.  However, inspite of the large literature which
now exists, a unified picture of multioscillator systems covering the various 
q-deformations and algebras that have been proposed, has not emerged.

The aim of the present work is to construct a general formalism which
may help one to develop such a unified picture.  We construct a theory
of generalized Fock spaces which has sufficient flexibility to encompass
all types of oscillator algebras that have been proposed in the past as
well as those that may be proposed in the future.   Using this
formalism, we are able to classify and clarify the interconnections that
exist between different q-deformations and different algebras.

In another paper [34] we have pointed out that as far as a single oscillator
is concerned, q-deformation does not lead to anything fundamentally new,
and is merely equivalent to a change of variable. A q-deformed
oscillator is just a different representation of the usual oscillator.
On the other hand, when we go to multioscillator systems, new things
are possible ; these are the new forms of statistics.  However, even
here, we find
that many of the q-deformations for multioscillator systems proposed in
the literature again belong to the category of substitution or change of
variables and should be regarded only as different representations 
of the well known Bose-Einstein or Fermi-Dirac statistics. 
Particular mention must be made of the work of
Greenberg [13,14] in this context. In contrast to most of the recent work
on multioscillator systems which are nothing but Bose or Fermi
statistics in disguise, Greenberg's proposal involves a new
statistics, called infinite statistics and this statistics is infact
based on a new Fock space which is much larger than the usual bosonic or
Fermionic Fock spaces.

In order to construct a unified theory, we find it convenient to start
with the underlying space of the allowed states of the system.  We
construct the creation and destruction operators $c^\dagger$ and c as outer
products of
the state vectors.  It is this device of starting with the set of state
vectors of the system as the primary concept, that cuts through the
jungle of different algebras which, {\it prima facie} look 
different, but on
closer examination are found to be related.  Two sets of creation and
destruction  operators which are related by substitution, operate on the
same space of  state vectors and hence describe essentially the same class
of systems.

We first formulate the theory of the
generalized Fock spaces. The key element is the notion of independence
of the permutation - ordered states. The largest linear vector space
constructed in this way is the super Fock space. The subsequent
specification of a subset of states in this space as null states leads
to many reduced Fock spaces. The general theory which applies to the
super Fock space as well as to the reduced Fock spaces, all of which are
to be called collectively as generalized Fock spaces, then allows us to
construct annihilation, creation and number operators. Whereas the
annihilation and creation operators and their algebras even for a
particular Fock space are not unique and many representations are
possible, a universal representation for the number operators valid for
all forms of statistics and algebras exists.

We apply the formalism to the super Fock space as well as to the bosonic
and fermionic Fock spaces, the latter being the most familiar examples
of reduced Fock spaces. Super Fock space is characterized by a unique
statistics named ``infinite statistics''. However, the same
infinite statistics can be represented by different algebras of $c$ and
$c^\dagger$ and we give a number of representions of inifnite statistics.
In the bosonic and fermionic Fock spaces, many different forms of
statistics can be defined and for each form of statistics many different
representations of $c$ and $c^\dagger$ are possible.

Although the main aim of the general framework presented here is to
develop a unified picture of the various forms of statistics and
algebras, the same formalism also allows us to construct a large number
of new forms of quantum statistics as well as new algebras of $c$ and
$c^\dagger$. In fact there is no limit to the number of possibilities.

An important part of our work deals with two-indexed systems. Our
general formalism for the single-indexed system applies to most
multi-indexed systems also since in most cases a multiplet of indices
can be mapped into a single index. But, we show that there exist certain
situations where such a mapping is not possible. Consequently, we
develop the general formalism for the two indexed system and discover
an enormously rich variety of novel forms of quantum statistics and
algebras.

The plan of the paper is as follows. The theory of the generalized Fock
spaces is given in Sec.2. In Sec.3 we apply the general theory to the
super Fock space and in Sec.4 we apply it to the reduced bosonic and 
fermionic Fock spaces.  Sec.4 also contains a new statistics called null
statistics. The alternative approach of starting with 
$cc^\dagger$ algebras and
deriving $cc$ relations therefrom is presented in Sec.5. Two-indexed
systems are treated in Sec.6 and Appendix. Sec.7 is devoted to summary
and discussion.

Since the paper is rather long, we may also point out that a quick 
overview of the paper may be obtained from the pictorial summaries given
in Figs.3 and 4 and the tabular information provided in Tables I-IV.

\newpage

\baselineskip=24pt
\setcounter{page}{5}

\noindent {\bf{2. Generalized Fock Spaces}}

Given a set of oscillators with indices g,h,i ${\ldots}$ m, we
construct the state vector
$$
\vert n_g, n_h \ldots n_m ;  1 \rangle \,=\, \vert \underbrace{1_g \cdots
1_g}_{n_g} \, \underbrace{1_h \cdots 1_h}_{n_h} \, \cdots \underbrace
{1_m \cdots 1_m}_{n_m} \, \rangle    \eqno(2.1)
$$
\noindent On the right-hand-side of this equation, $1_g, 1_h \ldots 1_m$
appear $n_g, n_h \ldots n_m$ times respectively and $n_g \ldots n_m$
denote the number of quanta with indices g ... m respectively.  Together
with the state (2.1) we consider the set of all states obtained through
all distinct permutations of the entries on the right-hand-side of (2.1).
Thus, we have
$$
\vert n_g, n_h \ldots n_m ; 2   \rangle \,=\, \vert \underbrace{1_g \cdots
1_g}_{n_g-1} \, 1_h 1_g \underbrace{1_h \cdots 1_h}_{n_h-1} \cdots
\underbrace{1_m \cdots 1_m}_{n_m}  \rangle   \eqno(2.2)
$$
$$
\vdots
$$
$$
\vert n_g, n_h \cdots n_m ; 35 \rangle \, = \, \vert 1_g 1_g 1_m 1_m \,
\underbrace{1_h \cdots 1_h}_{n_h} \ldots \underbrace{1_m \cdots
1_m}_{n_m-2} \rangle   \eqno(2.3)
$$
$$
\vdots
$$
$$
\vert n_g, n_h \cdots n_m ; s \rangle \, = \, \vert \underbrace{1_m
\cdots 1_m}_{n_m} \, \ldots \underbrace{1_h \cdots 1_h}_{n_h} \,
\underbrace{1_g \cdots 1_g}_{n_g} \, \rangle     \eqno(2.4)
$$
\noindent Here we have given examples of a few permutations.  In (2.2), 
the positions of one g-quantum and one h-quantum has been
interchanged and in (2.3), we have indicated a few more interchanges.
Collectively, we shall denote the set of all these states as
$$
\vert n_g, n_h \ldots n_m ; \mu \rangle \quad , \quad \mu = {
1,2} \ldots s      \eqno(2.5)
$$
\noindent where $s$ is the total number of distinct permutations and
$\mu$ labels each of these states.  It is easy to see that
$$
s = \frac{(n_g + n_h + \ldots n_m)!}{n_g!n_h! \ldots n_m!}    \eqno(2.6)
$$

\noindent (In eq.(2.3), $\mu$ has been put 35 rather arbitrarily).  We
assume the existence of a unique vacuum state corresponding to zero
occupation number for all the oscillators :
$$
\vert 0 \rangle \equiv \vert 0, 0, 0 \ldots 0 \rangle   \eqno(2.7)
$$

We shall first consider the set of all the s states given in (2.5) as
linearly independent.  Although they are linearly independent, they may
not be orthogonal to each other in general, nor are they normalized.
However, a state in one sector characterized by the occupation
numbers ${(n_g,n_h \ldots n_m)}$ is orthogonal to any state in
another sector characterized by another set of occupation numbers
${(n^{'}_g, n^{'}_h \ldots n^{'}_m)}$.  We can summarize these
statements by the equation :
$$
\langle n^{'}_g, n^{'}_h \ldots n^{'}_m ; \alpha \vert n_g,n_h \ldots n_m ;
\beta \rangle \, = \, \delta_{n^{'}_g n_g} \, \delta_{n^{'}_h n_h}
\ldots \delta_{n^{'}_m n_m} \, M_{\alpha \beta}     \eqno(2.8)
$$

\noindent Note in particular that the inner product vanishes even if 
a single occupation number does not match.  Within the same sector, 
the inner product is given by
$$
\langle n_g \ldots n_m ; \alpha \vert n_g \ldots n_m ; \beta \rangle
\,=\, M_{\alpha \beta}     \eqno(2.8a)
$$
where $M$ is a $s \times s$ hermitian matrix.  In fact, there is an infinite
set of matrices of varying dimensions, one corresponding to each sector
$\{n_g \ldots n_m\}$.  We choose all these matrices to be positive
definite.  This set of inner-product matrices M plays an important role
in the general formalism.

>From the set of linearly independent vectors given in (2.5), it is
possible to construct an orthonormal set of vectors which we shall
denote by a double-barred ket:
$$
\parallel n_g \ldots  n_m; \mu \gg ; \quad \mu = 1 . . . . . s    
\eqno(2.9)
$$
These satisfy the orthonormality relation 
$$
\ll n^{'}_g \ldots n^{'}_m ; \alpha \parallel n_g \ldots n_m ; \beta \gg \,
= \, \delta_{n^{'}_g n_g} \ldots \delta_{n^{'}_m n_m} \, 
\delta_{\alpha \beta}         \eqno(2.10)
$$

\noindent There is no unique way of doing this and the resulting 
orthonormal set is
not unique.  One may use Gram - Schmidt orthogonalization procedure or
calculate the eigenstates of the inner-product matrix $M$ or follow any
other method.  Whatever may be the method, one can write the relation
connecting the two sets of kets :
$$
\parallel n_g \ldots n_m ; \mu \gg \, =\, \sum_\nu X_{\nu \mu} \vert n_g
\ldots n_m ; \nu \rangle    \eqno(2.11)
$$
\noindent and the inverse relation :
$$
\vert n_g \ldots n_m ; \alpha \rangle \, = \, \sum_{\beta}
(X^{-1})_{\beta \alpha} \, \parallel n_g \ldots n_m ; \beta \gg
\eqno(2.12)
$$
\noindent where $X$ is a nonsingular matrix.  Although $X$ is not unique
(since it depends on the particular orthogonalization procedure used),
it is possible to show, using (2.8), (2.10), (2.11) and (2.12), 
that $XX^{\dagger}$ is the inverse of the innerproduct 
matrix\footnote{Eq(2.13) is analogous to the relation 
between the metric tensor and the vierbien in general relativity.}:
$$
M^{-1} \,=\, XX^{\dagger}     \eqno(2.13)
$$
\noindent Thus, one simple way of ensuring positivity of the
inner-product matrix is to choose a non-singular matrix $X$ and then
determine $M$ using (2.13).  Again it must be kept in mind that we are
dealing with an infinite set of matrices, $X$, one for each sector $\{ n_g
\ldots n_m \}$.  Also the orthonormality relation holds between vectors
from two different sectors and we have already used this in writing
(2.10).

The completeness relation for the orthonormal set of states 
{\hbox{$\parallel
n_g \ldots n_m ; \mu \gg $}} can be written in the form:
$$
I \,=\, \sum_{n_g \ldots n_m} \, \sum_{\mu} \, \parallel n_g \ldots n_m
; \mu \gg \, \ll n_g \ldots n_m ; \mu \parallel    \eqno(2.14)
$$
\noindent where I is the identity operator.  Substituting from (2.11) into
(2.14) and using (2.13), we get the resolution of the identity operator in
terms of the non-orthonormal set of states :
$$
I \,=\, \sum_{n_g \ldots n_m} \, \sum_{\lambda, \nu} \, \vert n_g \ldots
n_m ; \nu \rangle (M^{-1})_{\nu \lambda} \, \langle n_g \ldots n_m ;
\lambda \vert    \eqno(2.14a)
$$
\noindent It is convenient to define the projection operator
$$
P(n_g \ldots n_m) \,=\, \sum_{\mu} \parallel n_g \ldots n_m ; \mu \gg \,
\ll n_g \ldots n_m ; \mu \parallel     \eqno(2.15)
$$
$$
= \sum_{\lambda, \nu} \, \vert n_g \ldots n_m ; \nu \rangle
(M^{-1})_{\nu \lambda} \, \langle n_g \ldots n_m ; \lambda \vert
\eqno(2.16)
$$

\noindent so that we have
$$
I = \sum_{n_g \ldots n_m} \, P (n_g \ldots n_m)    \eqno(2.17)
$$
\noindent One can easily verify the following properties of the
projection operators :

$$
P (n_g \ldots n_m) \parallel n^{'}_g \ldots n^{'}_m ; \mu \gg \, =\,
\delta_{n_g n^{'}_g} \ldots \delta_{n_m n^{'}_m } \, \parallel n_g
\ldots n_m ; \mu \gg    \eqno(2.18)
$$
$$
P (n_g \ldots n_m) \vert n^{'}_g \ldots n^{'}_m ;\mu \rangle \, =\, 
\delta_{n_g n^{'}_g} \ldots \delta_{n_m n^{'}_m} \, \vert n_g \ldots n_m
; \mu \rangle      \eqno(2.19)
$$

\noindent It is worth noting that ${P (n_g \ldots n_m )}$
projects out any single state not only from the orthonormal set
$\parallel n_g \ldots n_m ; \mu \gg $ but also from the non-orthonormal
set $\vert n_g \ldots n_m ; \mu \rangle$.

In terms of the above projection operators it is very easy to construct
the number operators :
$$
N_k \, =\, \sum_{n_g \ldots n_k \ldots n_m} \quad n_k P(n_g \ldots n_k
\ldots n_m)       \eqno(2.20)
$$
\noindent which satisfy the following properties 
$$
N_k \parallel n_g \ldots n_k \ldots n_m ; \mu \gg \,=\, n_k \parallel
n_g \ldots n_k \ldots n_m ; \mu \gg      \eqno(2.21)
$$
$$
N_k \vert n_g \ldots n_k \ldots n_m ; \mu \rangle \, =\, n_k \vert n_g
\ldots n_k \ldots n_m ; \mu \rangle     \eqno(2.22)
$$
$$
[N_k, N_j] \, = \, 0  \quad   {\rm for \ any} \ k \ {\rm and} \ j    
\eqno(2.23)
$$

We now introduce the transition operators which connect states, lying in
different sectors.  Obviously it is enough to define the so called
annihilation and creation operators $c_j$ and $c^{\dagger}_j$.  We
define
$$
c^{\dagger}_j \,=\, \sum_{n_g \ldots n_j \ldots n_m} \, \sum_{\mu^{'}
\nu} \, A_{\mu^{'} \nu} \, \vert n_g \ldots (n_{j}+1) \ldots n_m ; \mu^{'}
\rangle \, \langle n_g \ldots n_j \dots n_m ; \nu \vert    \eqno(2.24)
$$
\noindent and $c_j$ as the hermitian conjugate of c$^{\dagger}_j$,
where $A_{\mu^{'} \nu}$ are a set of arbitrary (complex) numbers.  Note
that the span of $\mu^{'}$ is larger than that of $\nu$ and this is the
reason for the prime on $\mu$.
\noindent Specifically,

$$ 
\mu^{'} \,   =  \, 1 \ldots s^{'} \, ; \, s^{'} \,=\, \frac{(n_g + \ldots 
(n_{j}+1) + \ldots n_m) !}{n_g ! \ldots (n_j +1) ! \ldots n_m !} \eqno(2.25a)
$$
$$
\nu  \, =  1 \, \ldots s \, ; \, s = \frac{(n_g + \ldots n_j + \ldots n_m)!}{n_g!
\ldots n_j ! \ldots n_m!} \eqno(2.25b)
$$

\noindent Hence A is a rectangular matrix.  Since A is arbitrary in
general,  the relation(2.24) provides the most general definition of the creation
operator.  Even in this general case, it is possible to verify the
following commutation relation between the number operator defined in
(2.20) and the creation operator defined in (2.24)
$$
[c^{\dagger}_j, N_k ] \, =\, - c^{\dagger}_j \, \delta_{jk}    \eqno(2.26)
$$

\noindent The projection property given in (2.19) plays a crucial role in
the proof of (2.26).

So far we did not specify how the ordered state vectors $\vert n_g
\ldots n_m ; \mu \rangle $ are constructed.  In fact, in general there
is no need to specify any procedure for their explicit construction.
The formalism given so far holds whatever may be the explicit form of
their construction.  However, once annihilation and creation operators
c and $c^{\dagger}$ are introduced, there exists a natural procedure to
construct the set of ordered states using $c^{\dagger}$ and the vacuum
state $\vert 0 \rangle$.  This procedure has the advantage that the
arbitrariness of the matrix A introduced in (2.24)
disappears and A in fact gets determined in terms of M. Hence let us do
it.

In terms of creation operators, the ordered state (2.2) for instance is
constructed as follows :
$$
\vert \underbrace{1_g \cdots 1_g}_{n_g-1} \, 1_h 1_g \underbrace{1_h
\cdots 1_h}_{n_h-1} \ldots \underbrace{1_m \cdots 1_m}_{n_m} \rangle \,
= \, (c^{\dagger}_g)^{n_g-1} c^{\dagger}_h c^{\dagger}_g
(c^{\dagger}_h)^{n_h-1} \cdots (c^{\dagger}_m)^{n_m} \, \vert 0 \rangle
\eqno(2.27)
$$

\noindent and other states are constructed in a similar fashion.  We may
introduce the notation
$$
|n_g \ldots n_m ; \mu > \ = \ (c^{\dagger^{n_g}}_g \ldots 
c^{\dagger^{n_m}}_m ; \mu ) | 0 >     \eqno(2.28)
$$
$$
<n_g \ldots n_m ; \mu | \ = \ < 0 | (c^{n_m}_m \ldots 
c^{n_g}_g ; \mu )       \eqno(2.29)
$$

\noindent where $(c^{\dagger^{n_g}}_g \ldots c^{\dagger^{n_m}}_m ; \mu)$
is a permutation of the creation operators similar to the permutations
defined in eqn(2.1) - (2.5) and $(c^{n_m}_m \, \cdots c^{n_g}_g ; \mu)$
is defined 
as the hermitian conjugate of $(c^{\dagger^{n_g}}_g 
\ldots c^{\dagger^{n_m}}_m ; \mu)$. For
states constructed in this manner, there exists a simple formula
connecting states in ``adjacent'' sectors.  For instance, 
$$
\vert 1_j, \underbrace{1_g \cdots 1_g}_{n_g} \ldots \underbrace{ 
1_m \cdots 1_m}_{n_m} \rangle \, = \, c^{\dagger}_j \vert \underbrace{1_g \cdots 1_g}_{n_g}
\ldots \underbrace{1_m \cdots 1_m}_{n_m} \, \rangle   \eqno(2.30) 
$$

\noindent More generally, we may write
$$
c^{\dagger}_j \vert n_g \ldots n_j \ldots ; \lambda \rangle \,=\, \vert
1_j, n_g \ldots n_j \ldots ; \lambda \rangle      \eqno(2.31) 
$$
\noindent where $\vert 1_j, n_g \ldots n_j \ldots ; \lambda \rangle$ on
the right is a subset of states in which one `j'
quantum appears on the extreme left. Although the total number 
of states in the set $\vert n_g
\ldots (n_j + 1) \ldots ; \lambda^{'} \rangle$ is $s^{'}$ given by
(2.25a), the total number of states in the subset $\vert 1_j, n_g \ldots
n_j \ldots ; \lambda \rangle$ is s given by (2.25b).   Further the states
in the subset $\vert 1_j, n_g \ldots n_j ...; \lambda \rangle$ are given the
same ordinal number $\lambda$ as in the set 
$\vert n_g \ldots n_j \dots ; \lambda \rangle$.   This is possible 
since the quanta $\{ n_g \ldots n_j \ldots \}$ are permuted among 
themselves without disturbing the extra j-quantum
sitting on the extreme left.

We now substitute the expression for $c^{\dagger}_j$  given by 
(2.24) into the left hand side of (2.31).  We have 
$$ 
c^{\dagger}_j \vert n_g \ldots n_j \ldots ; \lambda \rangle 
$$
$$
= \, \sum_{n^{'}_g \ldots n^{'}_j\ldots} \, \sum_{\mu^{'} \nu} \, A_{\mu^{'}
\nu} \, \vert n^{'}_g \ldots (n^{'}_{j}+1) \ldots ; \mu^{'} \rangle \,
\langle n^{'}_g \ldots n^{'}_j \ldots ; \nu \vert n_g \ldots n_j 
\ldots ; \lambda \rangle
$$
$$
= \sum_{\mu^{'}\nu} \,  A_{\mu^{'} \nu} M_{\nu \lambda} \, \vert
n_g \ldots (n_{j}+1) \ldots  ; \mu^{'} \rangle     \eqno(2.32)    
$$
\noindent where we have used (2.8).  On comparing with (2.31), we get 
$$
\sum_{\nu} \, A_{\mu^{'} \nu} \, M_{\nu \lambda} \,=\, \delta_{\mu^{'}
\lambda}     \eqno(2.33) 
$$

\noindent From (2.33) we see that $\mu^{'} \,=\, \lambda$. 
This means that in (2.24),
only the subset $\vert 1_j, n_g \ldots n_j \ldots ; \lambda \rangle $
contributes and hence A is in fact a square matrix and is equal to the
inverse of M$^{-1}$:
$$A \,=\, M^{-1}     \eqno(2.34) 
$$
So, we may rewrite (2.24):
$$
c^{\dagger}_j \,=\, \sum_{n_g \ldots n_j \ldots} \, \sum_{\lambda, \nu}
\, (M^{-1})_{\lambda \nu} \, \vert 1_j, n_g \ldots n_j \ldots ; \lambda
\rangle \, \langle n_g \ldots n_j \ldots ; \nu \vert  \eqno(2.35) 
$$
\noindent Thus, A and hence c and c$^{\dagger}$ are completely determined in
terms of the set of M matrices.

Next, consider the expression for the number operator (eqs.(2.20),(2.15)
and (2.16)) :
$$
N_k = \sum_{n_g \ldots n_k \ldots n_m} n_k \sum_\mu \parallel n_g \ldots
n_k \ldots n_m ; \mu \gg \ll n_g \ldots n_k \ldots n_m ; \mu \parallel
\eqno(2.36)
$$
$$
= \sum_{n_g \ldots n_k \ldots n_m} n_k \sum_{\lambda \nu} 
\vert n_g \ldots n_k \ldots n_m ; \lambda > (M^{-1})_{\lambda\nu}
< n_g \ldots n_k \ldots n_m ; \nu \vert    \eqno(2.37)
$$

\noindent One may try to express this in terms of $c$ and $c^\dagger$.
Using eqs.(2.28) and (2.29),
$$
N_k = \sum_{n_g \ldots n_k \ldots n_m} n_k \sum_{\lambda \nu} 
(c^{\dagger^{n_g}}_g \ldots c^{\dagger^{n_k}}_k \ldots
c^{\dagger^{n_m}}_m ; \lambda) |0><0| (c^{n_m}_m \ldots c^{n_k}_k \ldots
c^{n_g}_g ; \nu ) (M^{-1})_{\lambda \nu}    \eqno(2.38)
$$

\noindent Apart from $c^\dagger$ and $c$, the above expression contains
the vacuum projector $|0><0|$. Later we shall give examples where $|0><0|$ is
determined as products of $c^\dagger$ and $c$ so that $N_k$ can be
expressed entirely in terms of $c^\dagger$ and $c$. However, in general,
this does not lead to simple results, whereas eq.(2.36) provides us with a
universal representation of number operators which is valid in all
cases.

So far, we regarded the set of state vectors $|n_g, n_h \ldots ; \mu > ;
\mu = 1 \ldots s$ (where $s$ is given by (2.6)) to be linearly 
independent and the resulting
generalized Fock space is the complete Fock space, which we shall call
the {\it super Fock space}.

We shall now show how to construct reduced Fock spaces. The motivation
for this is that many Fock spaces of physical interest such as the
bosonic Fock space or fermionic Fock space are reduced Fock spaces.
There are various ways of doing this. One may postulate relationships
between states connected by permutations, or one may disallow certain
permutations by equating them to null vectors. Yet another way to
achieve this is to use the permutation group $S_n$ acting on the
$n$-particle state. The super Fock space we have constructed consists of
all the representations of the permutation group. If we allow only
certain representations of $S_n$, we get a reduced Fock space.

All these possibilities are contained in the statement that in the space
of vectors $|n_g, n_h \ldots ;\mu>$, there are $r$ null vectors $(r <s)$
$$
\sum_{\mu} B^p_\mu |n_g,n_h \ldots ; \mu > = 0 \ ; \ p = 1,2, \ldots r ; r
<s.   \eqno(2.39)
$$

\noindent where $B^p_{\mu}$ are constants.  So, the dimension of the vector space in the sector $\{n_g, n_h \ldots
\}$ is reduced to $d$ given by
$$
d \ = \ s \ - \ r\,.    \eqno(2.40)
$$

An important class of reduced Fock spaces are those for which $d = 1$. Here, all
the states connected by permutation of indices will be taken to be
related to each other through equations of the type(2.39).  In
otherwords,the number of relations $r$ in eq. (2.39) is
$s-1$. We shall call this space as the {\it bosonic Fock space}. If we impose
the additional restriction : $n_g = 0$ or $1$ only, the resulting space
will be called {\it fermionic Fock space}.
This restriction can also be stated in the form of eq.(2.39) :
$$
|n_g, n_h \ldots ; \mu > \ = \ 0 \quad {\rm for} \quad n_g, n_h \ldots
\ge 2\,.    \eqno(2.41)
$$

We can define a new reduced Fock space, also of dimension $d=1$ in each
sector, by taking the set of all the permuted states as null states
except a single state (of a chosen order) which is taken to be the
allowed state. We shall call this as the Fock space of frozen order.

Another important class of reduced Fock spaces are those associated with
parastatistics [35-37], which we shall call parabosonic and parafermionic
Fock spaces. For these, the number of relations $r$ is smaller than for
bosonic or fermionic Fock spaces, so that the dimension $d$ satisfies
$$
1 < d < s \,. \eqno(2.42)
$$

{\it The formalism constructed in the present section is valid for all these
reduced Fock spaces also}, with the modification that all the summations
over $\mu,\nu$ etc. will now go over the range $1 \ldots d$ and
correspondingly, $X,M,$ and $A$ become $d\times d$ matrices.  There is
an arbitrariness in the choice of the d states.
Any choice of $d$ states will do, as long as they are non-null
states.

All these Fock spaces, the super Fock space as well as the reduced ones
will be collectively called generalized Fock spaces. To sum up this
discussion we may note that a generalized Fock space is completely
defined by stating what are the allowed states of the system.

We shall define statistics by the precise relationship linking states
obtained by permutation. In general, many relationships can be envisaged
and hence many different forms of statistics can reside within a particular reduced Fock space. However,
in super Fock space, all the states obtained by permutation are
independent and so there is a unique statistics associated with this
Fock space, namely ``infinite statistics''. Similarly,
in the Fock space of frozen order too, there exists only a single
statistics, named ``null statistics''.

In this section, we started with the generalized Fock space consisting
of the set of allowed states of the system and constructed the creation,
annihilation and number oeprators in terms of the outer products of state 
vectors. Do these $c$ and $c^\dagger$ form an operator algebra? 
In general, $c$ and
$c^\dagger$ constructed in this way may not form a simple algebra, or
even a closed algebra. Historically, it is the reverse route that has
been followed ; one postulates an algebra of $c$ and $c^\dagger$ and
then deduces the states allowed by the algebra. In this sense, a given
relation involving $c$ and $c^\dagger$ implicitly defines an inner
product and through it specifies the allowed and the null states of the
system. In practice, starting with an algebra is an easier procedure
and we shall use it in the later sections. Actually, it is
complementary to the approach described in this section. 

Therefore we have two equivalent ways of dealing with the generalized Fock
spaces. In the first approach, which we have formulated in this section
we construct $c$ and $c^\dagger$ in terms of the allowed states of the
system and the algebra of $c$ and $c^\dagger$ is then a derived
consequence. This is the more fundamental approach and is of universal
validity. In the second approach, which also we shall use in the latter
sections, we start with an algebra of $c$ and $c^\dagger$
and then determine the states of the system allowed by the $cc^\dagger$
algebra. Since the restrictions on the allowed states of the system can
generally be stated in the form of $cc$ relations, the first
approach can be characterized as $cc \rightarrow cc^\dagger$ while the
second is $cc^\dagger \rightarrow cc$. Within the second approach, 
we shall describe an elegant method to derive $cc$ relations from 
$cc^\dagger$ algebras.

For infinite statistics, there is no restriction on the allowed
states and $cc$ relations do not exist. Hence, for this statistics, 
the first approach should be interpreted as ``no $cc$
relation'' $\rightarrow cc^\dagger$ algebra while in the second approach,
starting with any particular $cc^\dagger$ algebra describing infinite
statistics one shows that there does not exist any $cc$ relation. In
those cases where the $cc^\dagger$ algebra depends on a continuous
parameter $q$, one can determine the values of $q$ where $cc$ relation
exists. Generally, these values of $q$ correspond to the boundary 
of the region in the parameter space where infinite statistics 
with positive definite $M$
exists. On this boundary, one or more eigenvalues of the $M$ matrices
become zero, thus leading to the emergence of the same number of null
vectors in the Fock space which can equivalently be interpreted as the
emergence of $cc$ relations. Thus the formalism unifies infinite
statistics residing on the super Fock space with the various forms of
statistics residing on reduced Fock spaces.

It must be noted that the inner product matrices $M$ occuring in
eq.(2.35) are quite arbitrary. Consequently, more than one realization or
representation of creation and destruction operators is possible. In
fact, it is this freedom to select arbitrary $M$ which enables one to
construct different algebras involving $c$ and $c^\dagger$, all
operating over the same Fock space.

To sum up, we construct a three-tiered structure consisting of Fock
space, statistics and algebra. Fock space is specified by the set of the
allowed states of the system. Statistics is defined by the nature  of the
symmetry of the allowed states under permutation. Algebra of the
creation and destruction operators is determined by the choice of the
inner product matrices $M$.

We shall construct different representations of
infinite statistics in Sec.3. In the bosonic and fermionic Fock spaces,
many forms of quantum statistics which include Bose and Fermi statistics
are possible. These and the null statistics in the Fock space of frozen
order will be taken up in Sec.4.

\newpage

\baselineskip=24pt
\setcounter{page}{19}

\noindent {\bf 3. Super Fock Space and Infinite Statistics}

\noindent {\bf 3.1 The standard representation $(M = 1)$}

Having taken all the s states to be independent, the simplest choice of
the matrix X is the unit matrix which implies M and A also to be unit
matrices (of appropriate dimensions) for all the sectors 
$\{ n_g, n_h \ldots\}$ :
$$
X = M = A = 1             \eqno(3.1)
$$
\noindent For this choice, which we shall call the standard
representation of infinite statistics, the ordered states (2.5) themselves
form an orthonormal set :
$$
\langle  n_g, n_h \ldots ; \mu \vert n_g, n_h \ldots ; \nu \rangle \,=\,
\delta_{\mu \nu}          \eqno(3.2)
$$

\noindent Further, the creation and annihilation operators are given by
$$
c^{\dagger}_j \,=\, \sum_{n_g, n_h\ldots} \, \sum_{\mu} \, 
\vert 1_{j} , n_g, n_h \ldots ; \mu 
\rangle \, \langle n_g, n_h \ldots ; \mu \vert 
\eqno(3.3) 
$$
$$
c_j \,=\, \sum_{n_g,n_h \ldots} \, \sum_{\mu} \vert n_g, n_h \ldots ;
\mu  \rangle \, \langle 1_{j} , n_g, n_h \ldots ; \mu \vert \eqno(3.4)
$$

\noindent From (3.3) and (3.4), using the orthogonality and completeness
relations, one can derive the cc$^\dagger$ algebra :
$$
c_i c^\dagger_j \, =\, \delta_{ij}      \eqno(3.5)
$$

\noindent This algebra was first proposed by Greenberg [13], in an
important paper on infinite statistics.

In the standard representation, we also have the useful identity :
$$
\sum_i c^\dagger_i c_i \ = 1 - |0><0| \eqno(3.6)
$$

\noindent Although this identity can be obtained from the eqs(3.3)
and (3.4) that define $c^\dagger$ and $c$, it is important to 
note that within Fock space it follows
from the $cc^\dagger$ algebra (eq.(3.5)). We shall prove it by showing
that eq.(3.6) is valid, when applied on any state in the super 
Fock space. First applying on $|0>$, we find that both sides are zero.
Next we apply on the other states $c^\dagger_k c^\dagger_\ell \ldots |0> $:
$$
\left\{ \sum_i c^\dagger_i c_i - 1 + |0><0| \right\} c^\dagger_j 
c^\dagger_k \ldots |0> = 0 \eqno(3.7)
$$

\noindent Using eq.(3.5) and noting $<0|c^\dagger_j \ = \ 0$
 we see that the left side of eq(3.7) infact vanishes, thus
completing the proof of eq(3.6).

In a different context, Cuntz [38] had studied the algebra defined by
eq(3.5) and by the relation :
$$
\sum_i c^\dagger_i c_i \ = \ 1\,. \eqno(3.8)
$$

\noindent Cuntz algebra is inconsistent with Fock
space, as can be seen by applying both sides of eq (3.8) on $|0>$.
In contrast, our eq(3.6), because of the inclusion of the vacuum 
projector term in it, is consistent with Fock space and is infact a
consequence of the algebra of $c$ and 
$c^\dagger$ in the standard representation of infinite statistics.

Putting $M=1$, the number operator $N_k$ given in eq(2.38) becomes
$$
N_k \ = \ \sum_{n_g..n_k..n_m} n_k \sum_\mu \left( c^{\dagger^{n_g}}_g
\ldots c^{\dagger^{n_k}}_k \ldots c^{\dagger^{n_m}}_m ; \mu\right)
|0><0| \left(c^{n_m}_m \ldots c^{n_k}_k \ldots c^{n_g}_g ; \mu
\right)\eqno(3.9) 
$$

\noindent Substitution of $|0><0|$ from (3.6) into this equation and a
straightforward but tedius calcualtion finally leads to 
$$
N_k \ = \ \sum_{n_g..n_k..n_m} \sum_\mu \left( c^{\dagger^{n_g}}_g
\ldots c^{\dagger^{n_k}}_k \ldots c^{\dagger^{n_m}}_m ; \mu\right)
c^\dagger_k c_k \left(c^{n_m}_m \ldots c^{n_k}_k \ldots c^{n_g}_g ; \mu
\right) \eqno(3.9a)
$$

\noindent This expression is identical to Greenberg's formula for $N_k$
[13], though it is written in a different form. This illustrates the
derivation of the representation of $N_k$ in terms of $c$ and
$c^\dagger$. As already mentioned, in general neither the procedure nor
the result is simple and in any case one does not need it. The universal
representation of $N_k$ given in eq(2.36) is sufficient.

\vspace{.5cm}

\noindent {\bf 3.2 $q$-mutators with real $q$}

Many other choices of M are possible. A particularly interesting choice is
the one in which M is given as a function of a real parameter q. Consider the inner product between the n-particle state
vectors with all occupation numbers unity : $\vert 1_g, 1_h \ldots ; \mu
\rangle $.  The inner product matrix M in this sector has dimension n !
$\times$ n! and its matrix element is taken to be
$$
\langle 1_g, 1_h \ldots ; \mu \vert \, 
1_g, 1_h \ldots ; \nu \rangle \, = q^J    \eqno(3.10)
$$
\noindent where q is a real number lying in the range $-1 < q < + 1$
and $J$ is the number of inversions required to transform the
state $\vert 1_g, 1_h \ldots ; \nu \rangle $ into the state $\vert 1_g,
1_h \ldots ; \mu \rangle$.  Number of inversions is the minimum number of
successive interchanges between adjacent quanta that will take the state
$\vert 1_g, 1_h \ldots ; \nu \rangle$ to $\vert 1_g, 1_h \ldots ; \mu
\rangle $.  For example,
$$
\langle 1_h 1_g 1_k \vert 1_g 1_h 1_k \rangle \,=\, q   \eqno(3.11)
$$
$$
\langle 1_h 1_k 1_g \vert 1_g 1_h 1_k \rangle \, =\, q^2   \eqno(3.12)
$$

The positivity of the q-dependent M matrices defined above has been
proved by Fivel [15] and Zagier [16].  In particular, Zagier  has given  the
explicit form of the determinant of the (n ! $\times$ n!) dimensional M
matrix for arbitrary n :
$$
det M \,=\, \prod^{n-1}_{k=1} \, [1-q^{k(k+1)}]^{(n-k)n!/k(k+1)}
\eqno(3.13)
$$
At $q=0$, $M$ is the unit matrix and so is positive-definite. Hence, it
will remain positive-definite in $-1 < q <1$, if det $M$ has no zeroes
there. According to Eq.(3.13), for real $q$, zeroes of det $M$ occur
only at $q = \pm 1$, thus proving the positive - definiteness of $M$ in
the range $-1 < q < 1$.

The inner product for states with occupation numbers larger than unity
(which are the same as states with repeated indices) is obtained from
the above inner product for states with distinct indices by symmetrizing
with respect to the repeated indices.  For example, consider
$$
\langle 2_g 1_m \vert 1_g 1_m 1_g \rangle \,=\, \langle 1_g 1_g 1_m
\vert 1_g 1_m 1_g \rangle     \eqno(3.14)
$$
\noindent Replace one of the g's by h in both the initial and final
states and thus get a matrix element with distinct indices which can be
calculated using(3.10).  This replacement can be done in (2!)$^2$ ways.
The sum of these (2!)$^2$ matrix elements divided by 2! is the required
answer.  Thus,
$$
\langle 1_g 1_g 1_m \vert 1_g 1_m 1_g \rangle \, =\, \frac{1}{2} \left[
\langle 1_g 1_h 1_m \vert 1_g 1_m 1_h \rangle  + 
\langle 1_g 1_h 1_m \vert 1_h 1_m 1_g \rangle  +  \right.
$$
$$
\left. \langle 1_h 1_g 1_m \vert 1_g 1_m 1_h \rangle  + 
\langle 1_h 1_g 1_m \vert 1_h 1_m 1_g \rangle  \right]   \eqno(3.15)
$$
$$
= q + q^2 \eqno(3.16)
$$
\noindent It is clear that the matrix M for repeated indices is obtained
from the higher dimensional matrix for distinct indices by a process of
collapse or reduction.  It can be shown that this process retains the
positivity of the matrix.

It is worth noting that the norm for $n$-particle state with all
the indices repeated is
$$
<n_g|n_g> = [n_g!]_q = [n_g]_q [n_g-1]_q \ldots [2]_q [1]_q \eqno(3.17)
$$

\noindent where the ``$q$-number'' $[n]_q$ is defined by  
$$
[n]_q \ \equiv \  \frac{1-q^n}{1-q} = 1 + q + q^2 + \ldots q^{n-1}.
\eqno(3.18)
$$

Inverting these $M$ matrices and defining the creation operator through
(2.35), one can derive the algebra of $c$ and $c^\dagger$. This
is more difficult than the reverse procedure which is the way this
subject really developed. Greenberg [14] proposed the $q$-mutator
algebra :
$$
c_i c^\dagger_j - q c^\dagger_j c_i \ = \ \delta_{ij} \eqno(3.19)
$$

\noindent The $q$-dependent $M$ matrices defined through (3.10), (3.15)
and (3.16) follow from this algebra. Fivel and Zagier then proved the
positivity of these $M$ matrices for $-1 < q < + 1$. Inspite of the fact
that we have not derived eq(3.19) from the $M$ matrices defined here, we
can assert its validity because the $M$ matrices completely determine
$c$ and $c^\dagger$. It is in order to emphasize this point that we have
presented the $M$ matrices first and the algebra of $c$ and $c^\dagger$
as a derived consequence.

\vspace{.5cm}

\noindent {\bf 3.3. New representations of infinite statistics}

First we briefly consider the $2$-parameter algebra
$$
c_ic^\dagger_j - q_1 c^\dagger_j c_i - q_2 \delta_{ij} \sum_k
c^\dagger_k c_k \ = \ \delta_{ij} \eqno(3.20)
$$

\noindent where $q_1$ and $q_2$ are real parameters. This may be
regarded as another representation of infinite statistics. Although the
determination of the full region in the $\{q_1,q_2\}$ parameter space
for which $M$ is positive definite is still an unsolved problem, one can
show [19] positivity of $M$ on the straight line defined by $q_1=0 ;
-1<q_2 < \infty$. Infact it is possible to map this whole line on to the
point $q_1=0 ; q_2=0$  by a redefinition of $c$ and $c^\dagger$ so that 
we just get back the standard representation defined by eq (3.5).

We next present two new algebras :
$$
c_i c^\dagger_j - c^\dagger_j c_i \ = \ \delta_{ij} p^{2\sum_{k<i}N_k}
p^{N_i} \eqno(3.21)
$$

\noindent and
$$ \left. \begin{array}{c}
c_i c^\dagger_j - p^{-1} c^\dagger_j c_i \ = 0 \quad {\rm for} \quad i \ne
j \\
\\
c_i c^\dagger_i - c^\dagger_i c_i \ = \ p^{N_i} \end{array} \right\} 
\eqno(3.22) 
$$

\noindent where $p$ is a real parameter and $N_i$ are number operators
already written down in Sec.2. Of these two algebras, the first one
(eq.3.21) is based on ordered indices, that
is, given any two indices $i$ and $j$, one must be considered larger or
smaller than the other. So, we may take the indices to be the natural
numbers $1,2,3\ldots$. 

Eqs (3.21) and (3.22) are again representations of the same infinite
statistics, for it is possible to map both these algebras on to
Greenberg's $q$-mutator algebra of eq (3.19) with the following
identification of the parameters: $p=q^{-1}$.  Temporarilly renaming the
$(c,c^\dagger)$ of eqs (3.21) and (3.22) by $(b,b^\dagger)$
and $(d,d^\dagger)$ respectively, the mapping or transformation
equations are
$$
b_i \ = \ p^{\sum_{k<i} N_k} p^{\frac{1}{2}N_i} c_i \eqno(3.23)
$$
$$
d_i \ = \ p^{\frac{1}{2}N_i} c_i\,. \eqno(3.24)
$$

\noindent Hence, it is clear that both the algebras of eqs (3.21) and
(3.22) lead to positive definite $M$ matrices for $p >+1$ and $p <-1$
(corresponding to $-1 < q < + 1$).

An interesting feature of the algebra (3.21) is the validity of ordinary
commutation relation between $c_i$ and $c^\dagger_j$ for $i \ne j$ as in
Bose statistics ; nevertheless the full algebra (3.21) describes
infinite statistics! 

In contrast to the situation for the $q$-mutator algebra of eq (3.19), where
the number operators can be expressed in terms of $c$ and $c^\dagger$
only after considerable algebraic manouvers [16,17] and the resulting
expressions are quite complicated, the corresponding expressions for the
algebras (3.21) and (3.22) are simple. For (3.21), we get 
$({\hbox{taking}} \ p>0)$
$$ \left. \begin{array}{lcl}
N_1 & = & {\displaystyle{\frac{1}{log \ p}}} \ log \ (c_1 c_1^\dagger - c^\dagger_1 c_1)
\\
& & \qquad \quad  \vdots  \\
N_i & = & {\displaystyle{\frac{1}{log \ p}}} \ log \ (c_i c_i^\dagger - c^\dagger_i c_i) - 2
\sum_{k<i} N_k \end{array} \right\} \eqno(3.25)
$$

\noindent For (3.22), the number operator is even simpler :
$$
N_i \ = \ \frac{1}{log \ p} \ log \ (c_i c_i^\dagger - c^\dagger_i c_i) 
\quad {\rm for \ all} \quad i\,. \eqno(3.26)
$$

\noindent Inspite of our ability to write down such formal expressions for
$N_i$ in terms of $c$'s and $c^\dagger $s, we must also point out that
they are not of much use. All that one ever needs of the number
operators are the properties contained in the eqs (2.21) - (2.23) and
(2.26) and as for explicit representation, eqs (2.36) and (2.37) will
do.

\vspace{.5cm}

\noindent {\bf 3.4. $q$-mutators with complex $q$}

We now generalize Greenberg's q-mutator algebra to complex q. This 
generalized algebra is based on ordered indices and is defined by the
following equations :
$$
c_i c^\dagger_j - q c^\dagger_j c_i = 0 \quad {\rm for} \quad i < j
\eqno(3.27)
$$
$$
c_i c^\dagger_i - p c^\dagger_i c_i = 1   \eqno(3.28)
$$
where q and p are complex and real parameters respectively and the
indices i,j etc. refer to any of the natural numbers 1,2,3, $\ldots$ .
The relation for the opposite order $i > j$ is derivable from (3.27) by
hermitian conjugation :
$$
c_i c^\dagger_j - q^\ast c^\dagger_j c_i = 0 \quad {\hbox{for}} 
\quad i > j   \eqno(3.27')
$$

\noindent and so it is not an independent relation.

Let us now calculate the inner product matrix for this algebra.  For
distinct indices, we find
$$
\langle 1_g 1_h 1_k \ldots ; \mu \vert  1_g 1_h 1_k \ldots ; \nu \rangle
\,=\, (q^\ast)^{{J}_+} \, q^{J_ -}     \eqno(3.29)
$$
\noindent where J$_+$ and J$_ {-}$ are the number of positive and
negative inversions  in the  permutation  $\nu \rightarrow \mu$.  The total
number of inversions (the sum of positive and negative inversions)
is the same as the number of inversions already defined  below eq.(3.10).
We further define an inversion as positive if it is a
transposition of indices from the ascending order to the descending
order and as negative if it is the reverse transposition.  For example,
(1,2) $\rightarrow$ (2,1) is a positive inversion while (2,1) $\rightarrow$
(1,2) is a negative inversion.  Thus we have
$$
\langle 1_3 1_1 1_2 \vert 1_2 1_1 1_3 \rangle \,=\, (q^\ast)^2 q
\eqno(3.29')
$$
\noindent since the permutation (213) $\rightarrow$ (312) contains two
positive and one negative inversions as shown in Fig.1.

The relationship between the algebra defined by eqs(3.27) and (3.28)
with complex $q$ (and $p = |q|$) and Greenberg's algebra defined by
eq.(3.19) with real $q$ can be given. Calling the creation operators for
the former and latter algebras as $c^\dagger(q)$ and $c^\dagger(|q|)$
respectively, the relationship is
$$
c^\dagger_i(q) \ = \ e^{i\theta \sum_{k<i} N_k} c^\dagger_i(|q|) 
\eqno(3.30)
$$

\noindent where $\theta$ is the phase of $q$ :
$$
q \ = \ |q| e^{i\theta} \eqno(3.31)
$$

\noindent and $N_k$ are the number operators defined in eq.(2.37).

As a consequence, the inner product matrices for the two algebras are
related by $\theta$-dependent unitary matrices:
$$
M(q) \ = \ T^\dagger(\theta) M (|q|) T(\theta) \eqno(3.32)
$$

\noindent where
$$
T^\dagger T \ = \ TT^\dagger \ = \ 1\,. \eqno(3.33)
$$

\noindent We have already given in eq.(3.13), Zagier's result 
for the determinant of $M$ for real $q$. For our algebra with complex
$q$, because of eq.(3.32), the determinant of $M$ for $n$ particles with
distinct indices is  
$$
{\rm det}  M_n (q) \,=\, \prod^{n-1}_{k=1} (1-\mid q 
\mid^{k(k+1)})^{\frac{(n-k)n!}{k(k+1)}}    \eqno(3.34)
$$
Zeroes of det $M_n(q)$ occur only on the circle $|q| = 1$.
Hence, $M_n(q)$ for complex $q$ is positive for $|q| <1$, and
we have thus extended the Fivel-Zagier result for positivity to complex
$q$ (for the case $p = |q|$).

In contrast to the situation for Greenberg's algebra, the inner product
matrix M for states with repeated indices cannot be derived from that
for states with distinct indices, for the present algebra.  For
n-particle states with all the indices repeated, the norm is the same as
in eqn(3.17) and (3.18), with $q$ replaced by $p$ :
$$
\langle n_g \mid n_g \rangle \,=\, (1+p) (1+p+p^2) \ldots
(1+p+p^2+\ldots p^{n_g -1})     \eqno(3.35)
$$

\noindent which is positive for p $> -1$.  For states with only some
indices repeated, M is a function of both q and p and the problem is
more complicated ; but we have verified positivity of the M's upto
3-particle states for $-1 < p < \mid q \mid^{-2}$.

If we choose $p = -1$, all states with repeated indices are forbidden.
This is just Pauli's exclusion principle.  Thus, the algebra defined by
eqs.(3.27) and (3.28) with $p = -1$ leads to {\it infinite statistics 
with exclusion principle}.

Infact, we can restrict both eqs(3.5) and (3.19) to $i \ne j$ and for
$i = j$ replace them by
$$
c_ic_i^\dagger + c^\dagger_i c_i = 1 \eqno(3.36)
$$

\noindent In all these cases, we have infinite statistics with
exclusion principle.

We give a compilation of the various algebras all representing infinite
statistics in Table I.

\noindent {\bf 3.5. Unitary transformations}
 
We must point out an important difference between the algebras
described by eqs(3.5), (3.19) and (3.20) on the one hand and the algebras
of eqs(3.21), (3.22), (3.27) and (3.28) on the other hand. It is easy to 
see that the former are covariant under the unitary transformations on 
the indices:
$$
c_k \ \rightarrow \ \sum_m \ U_{km} \ c_m \ ; \ U^\dagger U \ 
= \ UU^\dagger \ = \ 1\,. \eqno(3.37)
$$

\noindent In fact, one can show [19] that under certain
conditions eq.(3.20) is the most general bilinear algebra of $c_i$ and
$c^\dagger_j$ invariant under this unitary transformation.
This property is violated for the complex $q$-algebra of eqs(3.27) and
(3.28) as well as the algebras of eqs (3.21) and (3.22). Covariance 
under the unitary
transformation is desirable in a general context since it is closely
connected to the superposition principle in quantum mechanics [3].
Nevertheless, algebras violating this requirement have been proposed in
the recent literature, either because of the possibility of applications
to specific systems in condensed matter physics or because of
mathematical motivation. Hence, we include such algebras in our
investigation.

\noindent {\bf 3.6 Thermodynamics}

A detailed treatment of thermodynamic aspects is outside the scope of
the present paper. Nevertheless a brief statement is in order. 

Consider the partition function for the canonical ensemble :
$$
Z \ = \ Tr e^{-\beta H} \eqno(3.38)
$$

\noindent where $\beta = (kT)^{-1}$ and $H$ is the Hamiltonian of the
system. If the system consists of noninteracting particles, we have
$$
H \ = \ \sum_i \epsilon_i N_i \eqno(3.39)
$$

\noindent where $N_i$ are the number operators and $\epsilon_i$ are the
single-particle energies. Hence, using the orthonormal set $\|\ldots n_i
\ldots ; \mu \gg$ in the evaluation of the trace in eq.(3.38), we get
$$
Z \ = \ \sum_{n_1,n_2\ldots} d(\ldots n_j \ldots ) 
e^{{\displaystyle{-\beta \sum_i \epsilon_i n_i}}} \eqno(3.40)
$$

\noindent where $d(\ldots n_j \ldots )$ is the dimension of the sector
$\{\ldots n_j \ldots \}$, defined in \linebreak eq.(2.40), and the summation is
over all the allowed occupation numbers $n_i$ for all $i$.

All the thermodynamics of the system can be derived from the partition
function. Further, the set of $d(\ldots n_j \ldots)$ on the right hand
side of eq.(3.40) provide invariant characteristics of the Fock space
that are not dependent on the particular statistics or algebra living in
that Fock space. Hence, thermodynamics is the same for all forms of
statistics and algebras residing in a given Fock space.

For super Fock space and the associated infinite statistics, for which
$d$ is equal to $s$ defined by eq.(2.6) : 
$$
d \ = \ n!/\Pi_i n_i! \quad ; \quad n \ = \ \Sigma_i n_i\,, \eqno(3.41)
$$
the thermodynamics is independent of the particular representation or
algebra of $c$ and $c^\dagger$. In particular, if one uses the
$q$-mutator algebra (eq.(3.19)) to describe infinite statistics, the
properties of the system in thermodynamic equilibrium are independent of
$q$, provided $q$ lies in the range $-1 < q < 1$. Greenberg [14] had
originally envisaged small violations of Fermi or Bose statistics by
choosing $q=\mp (1-\epsilon), \epsilon$ small and positive. We see that
equilibrium thermodynamics will not manifest such small violations.

The expression for $d$ given by eq.(3.41) is the same as Boltzmann
counting and hence Greenberg has called infinite statistics as ``quantum
Boltzmann statistics''. However, since the Gibbs factor $1/n!$ is
missing, the statistical mechanics of a system of free particles obeying
infinite statistics with the index $i$ in (3.40) interpreted as momentum
will suffer Gibbs paradox [20]. Therefore, the super Fock space and the
associated infinite statistics cannot be naively applied to familiar
physical systems. In our approach, super Fock space plays only the role
of a mathematical template which can be used for carving out various
physical systems.

As long as one takes the Hamiltonian to be of the form (3.39) which is
the only correct form for noninteracting particles, the statistical
mechanics and thermodynamics of the system are independent of the
algebra of $c$ and $c^\dagger$. One can then construct interaction terms
involving $c$ and $c^\dagger$ as in the usual many-body theory and their
effect will of course depend on the algebra.

Recent literature contains many calculations on the statistical
distributions or other thermodynamic quantities for the Hamiltonian
$$
H \ = \ \sum_i \epsilon_i c^\dagger_i c_i \eqno(3.42)
$$

\noindent with $c$ and $c^\dagger$ satisfying some $q$-deformed
algebra. Since $c^\dagger_i c_i$ is not a number operator in general,
the physical meaning of the Hamiltonian (3.42) is not a priori clear,
although one may suppose that it takes into account of some type of
interactions.

\newpage


\baselineskip=24pt
\setcounter{page}{33}

\noindent {\bf{4. Statistics and Algebras in Bosonic, 
Fermionic and Frozen Fock Spaces}}

We introduce two types of reduced Fock spaces in sub section 4.1.  In one,
we take any two states obtained by permutation to be related to each
other and we allow all occupation numbers :  $n_i = 0,1,2, \ldots$ for
all $i$. In the other, any two states obtained by permutation are again
related to each other but occupation numbers are restricted by exclusion
principle : $ n_i = 0$ and 1 only.  We shall call the former as bosonic
Fock space and the latter as fermionic Fock space.

In subsection 4.2, we introduce the Fock
space of frozen order, in which permutations are forbidden and 
this also comes in two varieties, bosonic and fermionic.

\vspace{.5cm}

\noindent {\bf 4.1 Bosonic and Fermionic Fock Spaces}

A consistent way of defining a general relationship among permuted
states for both the bosonic and fermionic Fock spaces is to take 
$$
\vert n_1, n_2 \ldots ; \mu \rangle \,=\, q^J  \vert n_1, n_2 \ldots ;
1 \rangle \eqno(4.1)
$$
\noindent where $q$ is a complex number and $J$ is the number of
inversions in the permutation : $1 \rightarrow \mu$. The number of
inversions was already defined in Sec.3.  Although $J$ was defined there
for the case in which the occupation numbers were restricted to unity,
the same definition can be extended to larger occupation numbers. 
We assume $q \ne 0$. The case $q=0$ will be considered separately.
Many different forms of statistics as well as various algebras of 
$c$ and $c^\dagger$ can be shown to be contained as particular cases of
eq.(4.1).

We first make some general remarks amplifying the meaning of eq(4.1).
This equation can be obtained by repeated application of the elementary
relation : 
$$
| \ldots 1_i \ 1_j \ldots > \ = \ q | \ldots 1_j 1_i \ldots >\,, \quad
{\rm for} \quad i > j \eqno(4.2)
$$

\noindent where, except for the two adjacent quanta of indices $i$ and
$j$ which are interchanged, all other quanta are left unchanged. In
eq.(4.2), we have used the same notation for the state vectors as on the
right hand side of eqs(2.1)-(2.4). It is clear that, for any $q$ other
than $\pm 1$, eq.(4.2) makes sense only if the indices $i$ and $j$ are
ordered through an inequality (to be specific, $i > j$). Hence we have
taken the indices to be the natural numbers $1,2,3\ldots$in writing
eq.(4.1).

The relationship among the state vectors given in eq(4.1) or (4.2) can
be equivalently expressed as a quadratic relation between two creation
operators or two annihilation operators. For any state $|\ldots >$,
using eq.(2.30) or (2.31) we have
$$
c^\dagger_i c^\dagger_j | \ldots > \ = \ | 1_i 1_j \ldots >
$$
$$
c^\dagger_j c^\dagger_i | \ldots > \ = \ | 1_j 1_i \ldots >
$$

\noindent Comparing with eq.(4.2), we get
$$
c^\dagger_i c_j^\dagger | \ldots > \ = \ q c_j^\dagger c^\dagger_i |
\ldots >\,, \quad  {\rm for } \quad i > j 
$$

\noindent Since this is valid for any state $| \ldots >$, we can write
$$
c_i^\dagger  c_j^\dagger - q c_j^\dagger  c_i^\dagger \, = 0 \quad 
{\rm for}  \quad i > j    
$$

\noindent or, 
$$
c_j  c_i - q^\star c_i c_j \ = \ 0 \quad 
{\rm for}  \quad i > j.   \eqno(4.3)
$$

The above $cc$  relation (eq 4.3) is common to both bosonic and
fermionic spaces. But for the fermionic space there exists the additonal
restriction :
$$
c^\dagger_i c^\dagger_i \ = \ 0\,, \quad {\rm or} \quad c_i c_i \ = \ 0
\eqno(4.4)
$$

All the states $|n_1,n_2 \ldots ; \mu>$ for fixed occupation numbers
$(n_1,n_2 \ldots)$ but different values of $\mu$ being related to each
other through eq.(4.1), the dimension of the vector space in any sector
$(n_1, n_2 \ldots)$ is reduced to unity. So, for each sector we choose
one standard vector $|n_1,n_2 \ldots ; 1>$ which we rewrite as $|n_1,n_2
\ldots>$, dispensing with $\mu$ completely.

The matrices $X$ and $M$ then become numbers related by
$$
M^{-1} \ = \ X X^\star\,. \eqno(4.5)
$$

\noindent Eq.(2.11) becomes
$$
\parallel n_1, n_2 \ldots \gg \,=\, X(n_1, n_2 \ldots) \vert n_1, n_2
\ldots \rangle    \eqno(4.6)
$$
\noindent and we have
$$
\ll n_1, n_2 \ldots \parallel n^{'}_1, n^{'}_2 \ldots \gg \,=\, 
\delta_{n_1 n^{'}_1} \, \delta_{n_2 n^{'}_2} \ldots   \eqno(4.7)
$$
\noindent From eq.(2.35), we get
$$
\begin{array}{lll}
c^\dagger_j & = & {\displaystyle{\sum_{n_1 \ldots n_j\ldots}}} \, 
\vert 1_j, n_1, n_2 \ldots
n_j .. \rangle \, \langle n_1, n_2 \ldots n_j .. \vert \, M^{-1} (n_1,
n_2 ..) \\
& \\
&=& {\displaystyle{\sum_{n_1 \ldots n_j\ldots}}} \, q^{\sum_{i<j} n_i}  \,
\vert n_1, n_2 \ldots
(n_{j}+1) .. \rangle \, \langle n_1, n_2 \ldots n_j .. \vert \, 
M^{-1} (n_1, n_2 \ldots) \end{array}
$$

\noindent where we have pushed the $j$ quantum to the right of all $i$
quanta for $i < j$ using eq.(4.2). Also using eqs.(4.5) and (4.6),
$$
\begin{array}{lll}
c^\dagger_j = & {\displaystyle{\sum_{n_1 \ldots n_j}}} \, 
q^{\sum_{i<j} n_i}  \,{\displaystyle{ \frac{X(n_1 .. n_j
..)}{X(n_1 .. (n_{j}+1)..)}}} \, \parallel n_1, n_2 \ldots (n_{j}+1) .. \gg \,
\ll n_1, n_2 \ldots n_j .. \parallel    
 \end{array}       \eqno(4.8)
$$
\noindent Eq.(4.8) and its hermitian conjugate define the most 
general form of creation and
destruction operators for the bosonic and fermionic Fock spaces.

If we assume factorization of the norm M as well as that of X :
$$
X(n_1, n_2 \ldots) \,=\, \phi (n_1) \phi (n_2) \ldots,   \eqno(4.9)
$$
\noindent then we get
$$
c^{\dagger}_j = \sum_{n_1 \ldots n_j \ldots} \, q^{\sum_{i < j}n_i} \,
\frac{\phi(n_j)}{\phi(n_j+1)} \, \parallel n_1 .. n_j +1 .. \gg \, \ll
n_1 .. n_j ..  \parallel     \eqno(4.10)
$$
$$
c_j = \sum_{n_1 \ldots n_j \ldots} \, (q^{\ast})^{\sum_{i < j}n_i} \,
\frac{\phi^\ast(n_j)}{\phi^\ast(n_j+1)} \, \parallel n_1 .. n_j .. \gg \, \ll
n_1 .. (n_j+1)  ..  \parallel     \eqno(4.11)
$$

We now construct the operator algebras for c$^{\dagger}$ and c given by
eqs.(4.10) and (4.11). First of all, {\it for any form of $\phi(n_i)$}, we
get 
$$
\left. \begin{array}{llr}
c_i c^{\dagger}_j - q c^{\dagger}_j c_i & = & 0  \quad for \quad i < j \\
& \\
c_i c^{\dagger}_j - q^\ast c^{\dagger}_j c_i & = & 0  \quad for \quad i > j
\end{array} \right\}    \eqno(4.12)
$$

\noindent For i = j, the algebra depends on the choice of q and
$\phi(n_i)$. We find
$$
c_j c_j^\dagger \quad = \quad |q|^{2 \sum_{i<j} N_i} \left\vert
\frac{\phi(N_j)}{\phi(N_j+1)} \right\vert^2 \eqno(4.13) 
$$
$$
c_j^\dagger c_j \quad = \quad  |q|^{2 \sum_{i<j} N_i} \left\vert
\frac{\phi(N_j-1)}{\phi(N_j)} \right\vert^2 \eqno(4.14) 
$$

\noindent where $N_i, N_j \ldots$ are number operators and $\phi(N_i)$ 
is the function introduced in
eq.(4.9). So, we can write the generalized commutation relation
$$ \left. \begin{array}{c}
c_j c_j^\dagger - p c^\dagger_j c_j \ = \ |q|^{2 \sum_{i<j} N_i} f(N_j)
\\
\\
f(N_j) \ = \ \left \vert \frac{\phi (N_j)}{\phi(N_j+1)} \right \vert^2
-p \left \vert \frac{\phi(N_j-1)}{\phi(N_j)} \right \vert^2 \end{array}
\right\} \eqno(4.15)
$$

\noindent where $p$ is any (complex) number. Eqs.(4.12) and (4.15) constitute
the $cc^\dagger$ algebra for $q$-statistics defined by the $cc$ algebra
of eq.(4.3).

What we have derived above can be regarded as the most general algebra
of creation and annihilation operators for $q$-statistics, subject to
the assumption of the factorization of the normalization factor implied
by eq.(4.9). In particular, since the function $\phi(N_j)$ occuring in
eq.(4.15) is arbitrary, we may regard our equations as constituting an
infinite-parameter deformation of the algebra of $c$ and $c^\dagger$.
Exploiting the arbitrariness of $\phi(N_j)$ we can get many simpler
forms of the algebra which we now describe. 

\noindent {\bf (i)} Choosing $\phi(n_j)$ to be a constant independent of $n_j$
but noting that $\phi(-1)$ must vanish\footnote{On applying both sides
of eq.(4.14) on the vacuum state, the left hand side vanishes and so for
consistency, $\phi(-1)=0$}, eq.(4.15) becomes
$$
c_j c^\dagger_j - p c^\dagger_j c_j \ = \ |q|^{2 \sum_{i<j} N_i} \left\{
1-p(1-\delta_{N_j,0}) \right\} \eqno(4.16)
$$
\noindent where $\delta_{N_j,0}$ is the Kronecker delta, which can also be
represented as $sin \ \pi N_j / \pi N_j$.

\vspace{.5cm}

\noindent{\bf (ii)} Putting $p=0$ in eq.(4.16) we get the simple algebra :
$$
c_j c^\dagger_j \ = \ |q|^{2 \sum_{i<j} N_i} \eqno(4.17)
$$

\vspace{.5cm}

\noindent{(iii)} As a third possibility, we may choose $f(N_j)$ in eq.(4.15)
to be unity and $p$ real so that we have the algebra :
$$
c_j c^{\dagger}_j - p  c^{\dagger}_j c_j \, = \, |q|^{2 \sum_{i<j} N_i}     
\eqno(4.18)
$$

\noindent This choice requires that the function $\phi(n)$ must satisfy
the equation
$$
\left\vert \frac{\phi(n_j)}{\phi(n_j+1)} \right \vert^2 \ - \ p
\left\vert \frac{\phi(n_j-1)}{\phi(n_j)} \right\vert^2 \ = \ 1 \eqno(4.19)
$$

\noindent The solution of this functional equation (with the constraints
that $\phi(0)$ is finite and $\phi(-1)$ vanishes) is
$$
|\phi(n)|^{-2} \ = \ [(n)!]_p \equiv [n]_p [n-1]_p \ldots [1]_p
\eqno(4.20)
$$

\noindent where
$$
[n]_p \ = \ \frac{p^n-1}{p-1} \ = \ 1+p+p^2+\ldots p^{n-1} \eqno(4.21)
$$

\noindent For $p=1$, the ``p-numbers'' in
eqs(4.20) and (4.21) become ordinary numbers.

\vspace{.5cm}

\noindent {\bf (iv)} As a final choice we put $p=|q|^2$ in eqs (4.18) - (4.21).
As a consequence of this choice, the right hand side of eq (4.18)
becomes a bilinear function of $c$ and $c^\dagger$. For, from eq.(4.14),
we can prove an identity :
$$ 
(|q|^2-1) \sum_{k<j} c^\dagger_k c_k  =  \sum_{k<j} |q|^{2 \sum_{i<k}
N_i} \left \vert \frac{\phi(N_k-1)}{\phi(N_k)}\right \vert^2 (|q|^2-1)
$$
$$
=  \sum_{k<j} |q|^{2 \sum_{i<k} N_i} (|q|^{2N_k}-1) 
$$
$$
=  |q|^{2 \sum_{i<j} N_i} -1  \eqno(4.22)
$$	

\noindent where we have used eqs(4.20) and (4.21) with $p=|q|^2$. Hence
eq.(4.18) can be replaced by
$$
c_jc^\dagger_j - |q|^2 c^\dagger_j c_j \ = \ 1 + (|q|^2-1) \sum_{k<j}
c^\dagger_k c_k \eqno(4.23)
$$

Any one of the eqs(4.16), (4.17), (4.18) or (4.23) together with
eq.(4.12) constitute a $cc^\dagger$ algebra and one can construct many
more examples of $cc^\dagger$ algebras, all of which correspond to the
same $q$-statistics specified by eq.(4.3). We have so far assumed that
the occupation numbers are unrestricted : $n_j \ge 0$ and so these
algebras are valid for the bosonic $q$-statistics, namely $q$-statistics
in the bosonic space.

The fermionic $q$-statistics is defined by eqs(4.3) and (4.4). In this
case, $n_j$ in eqs(4.10) and (4.11) can take the value $0$ only, while
all the other occupation numbers will range over $0$ and $1$. So, the
only arbitrary parameter that enters the definition of $c$ and
$c^\dagger$ is $r = \phi(0)/\phi(1)$. The $cc^\dagger$ relation for $i
\ne j$ is still the same as in eq(4.12) while the $c_j c^\dagger_j$
relation of eq(4.15) can be simplified to
$$
c_jc^\dagger_j - p c^\dagger_j c_j \ = \ |q|^{2 \sum_{i<j} N_i} |r|^2
\left\{ \delta_{N_j,0} - p \delta_{N_j,1} \right \} \eqno(4.24)
$$

\noindent Further we can generally prove the fermionic analogue of 
the identity which was earlier proved in eq.(4.22) for the bososnic 
case only for a special choice of $\phi(n)$:
$$ \begin{array}{lcl}
(|q|^2-1) \sum_{k<j} c^\dagger_k c_k & = & (|q|^2-1) \sum_{k<j} 
|q|^{2 \sum_{i<k} N_i} |r|^2 \delta_{N_k,1} \\
\\
& = & \sum_{k<j} |q|^{2 \sum_{i<k} N_i} (|q|^{2N_k}-1) |r|^2 \\
\\
& = & (|q|^{2 \sum_{i<j} N_i} -1)|r|^2 \end{array} \eqno(4.25)
$$	

\noindent Hence, eq (4.24) can be rewritten as
$$
c_jc^\dagger_j - p c^\dagger_j c_j \ = \ \{ |r|^2 + (|q|^2 - 1)
\sum_{k<j} c^\dagger_k c_k\} (\delta_{N_j,0} - p \delta_{N_j,1})
\eqno(4.26)
$$

\noindent Eqs (4.12) and (4.26) constitute the general algebra for the
fermionic $q$-statistics. In contrast to the eq(4.15) of the bosonic
case, the fermionic case does not have the freedom of infinite - parameter 
deformation.

On the righthand side of eq.(4.26), apart from the curly bracket \{~~\}
which is bilinear in $c$ and $c^\dagger$, there are kronecker deltas
which depend on the operator $N_j$. A simpler relation is obtained for
the choice $p = -1$, since $\delta_{N_j,0} + \delta_{N_j,1} \ = \ 1$. We
then have
$$
c_jc^\dagger_j + c^\dagger_j c_j \ = \  |r|^2 + (|q|^2 -1) 
\sum_{k<j} c^\dagger_k c_k \eqno(4.27)
$$

\noindent A further choice $|r|^2 = 1$ gives
$$
c_jc^\dagger_j + c^\dagger_j c_j \ = \ 1 + (|q|^2 -1) 
\sum_{k<j} c^\dagger_k c_k \eqno(4.28)
$$

\noindent which is the fermionic analogue of eq.(4.23).

The algebra given by eqs(4.3), (4.12) and (4.23) (for real $q$) are
covariant under {\it the quantum group} $SU_q(n)$ where $n$ is the total
number of indices. This is true of the fermionic algebra of eqs(4.3),
(4.12) and (4.28) also. See ref [21-31] for this relationship to the
theory of quantum groups. However, our approach based on states related
by $q$-statistics (eq(4.1)) does not involve any notion of quantum group
{\it per se} and so we shall not describe it. The formalism of the 
generalized Fock space appears to be
capable of incorporating many general algebraic structures. In
particular, the $c_jc_j^\dagger$ algebra of eq(4.15) constitutes a
general infinite-parameter deformation while the quantum-group related
algebra of eq(4.23) is only a particular case of this.  

We have constructed the general algebras for arbitrary complex $q$.
The choice of $q$ defines the symmetry of the state under permutation.
Hence we call the different choices of $q$ as different forms of
statistics. Particular values of $q$ such as $q = \pm 1$ or 
$q = e^{i\theta}$ are of
special interest, although these cases are all contained in the general
formulae already given. We shall call the statistics obtained for $q =
+1, -1$ and $e^{i\theta}$ as Bose statistics, Fermi statistics or
``fractional'' statistics respectively. Either the bosonic or fermionic
Fock space can be used to construct any one of the various forms of
statistics. Thus, within our
terminology, it is perfectly possible for instance to have Fermi
statistics residing in bosonic Fock space or vice versa.

Within a particular statistics, different choices of $M$ or $\phi$
correspond to different representations of $c_j$ and $c^\dagger_j$,
which in turn lead to different $cc^\dagger$ algebras. If we assume the
factorization given in eq(4.9), for a given $q$, only the $i=j$ part of
the $cc^\dagger$ algebra depends on the representation.

For the sake of clarity, all these forms of statistics and algebras are
exhibited in Tables II and III for the bosonic and fermionic spaces
respectively. 

Several comments are in order, concerning the contents of these tables.
The algebra
$$
c_ic_j - c_j c_i = 0, \quad {\rm for} \quad i \ne j \eqno(4.29)
$$
$$
c_ic^\dagger_j - c^\dagger_j c_i = 0, \quad {\rm for} \quad i \ne j 
\eqno(4.30)
$$
$$
c_j c^\dagger_j \ = \ 1 \eqno(4.31)
$$

\noindent provide the simplest representation for Bose statistics based on
the simple choice of $M$ or $\phi$ as unity. This is analogous to the
standard representation (eq(3.5)) for infinite statistics but very
different from it. In fact they live in entirely different Fock spaces.
In recent literature, eq(4.31) has been sometimes confused with
infinite statistics. Eqs.(4.31) coupled with eqs(4.29) and (4.30) infact
describes Bose statistics only, although in a noncanonical
representation. 

Attention may be drawn to the ``anticommuting
bosons'' and the ``commuting fermions'' shown in Tables II and III
respectively.  The ``commuting fermions'' have been called hard-core
bosons in condensed-matter-physics literature.  Our terminology seems
more appropriate since they are really fermions in disguise, living in
the fermionic Fock space. 

The precise connection of our ``fractional'' statistics to the exchange
property of the wavefunctions proposed [39,40] for one and
two-dimensional systems needs further study. In particular, a suitable
mapping of the ordered indices to coordinates in one and two dimensions
is necessary. 

Among all the different algebras given in Tables II and III, only two of
them, namely the canonical representations for the bosons and fermions
have the distinction of being covariant under the unitary
transformations that mix the indices (see eq.(3.37)).  
All the other algebras violate this requirement, although, as we 
have already mentioned, two of them, one in the bosonic and the 
other in the fermionic Fock spaces, are covariant under the quantum 
groups SU$_q$(n).

The examples contained in these tables can be called deformed-oscillator
algebras on which many papers [1-12,21-27,32,41-42] have been written
recently. Inspite of the multiplicity of these deformed algebras,  we
must not ignore the basic fact that {\it they are all avatars of just two
primary constructions which may be taken to be the canonical bosonic
algebra for the bosonic Fock space and the canonical fermionic algebra for
the fermionic Fock space}.   All the different forms of statistics
belonging to the bosonic Fock space as well as their various algebraic
representations are related to each other  and to the canonical bosonic
algebra and similar is the situation
for the fermionic Fock space. Given two different forms of statistics
within the bosonic Fock space characterized by the statistics parameters
$q_1$ and $q_2$ respectively, or/and two different representations
characterized by the functions $\phi_1(n)$ and $\phi_2(n)$ respectively,
it is easy to get from (4.10) the relationship
$$
c^\dagger_j(2) \ = \ \left( \frac{q_2}{q_1}\right)^{\sum_{i<j} N_i} 
\frac{\phi_2(N_j-1)}{\phi_2(N_j)} \frac{\phi_1(N_j)}{\phi_1(N_j-1)}
c^\dagger_j(1) \eqno(4.32)
$$

\noindent The corresponding equation for the fermionic Fock space is
$$
c^\dagger_j(2) \ = \ \left( \frac{q_2}{q_1}\right)^{\sum_{i<j} N_i} \left(
\frac{r_2}{r_1}\right) c^\dagger_j(1) \eqno(4.33)
$$

\noindent Hence, from the point of view of Fock space, there is nothing 
new in all these deformed oscillators.

In the above equations, the number operator $N_i$ are given by
$$
N_i \ = \ \sum_{n_1 \ldots n_i \ldots} n_i \vert\vert n_1 \ldots n_i
\ldots \gg \ll n_1 \ldots n_i \ldots \vert\vert  \eqno(4.34)
$$

\noindent and, written in this fashion in terms of the normalized states
$\vert\vert \gg$, {\it $N_i$ is the same in all the different forms of
statistics (within the bosonic or fermonic Fock spaces) and in all 
the different representations}.

Using eq.(4.14) it is also possible to express the number
operators in terms of c and c$^{\dagger}$ in any statistics or any
representation within the bosonic or fermionic Fock space.  
Noting that for the canonical bosonic algebra 
$$
q = 1 \ ; \ \phi(n) = \frac{1}{(n!)^{1/2}} \quad ,         \eqno(4.35)
$$

\noindent we get from eq.(4.14)
$$
N_i = c^{\dagger}_i c_i.  \eqno(4.36)
$$

\noindent For other statistics and representations also, the set of 
equations obtained from eq.(4.14) starting with j $=$ 1 and
successively increasing j, can be seen to be implicit recursion formulae
for all $N_j$. Thus we have
$$ \left. \begin{array}{lcl}
{\displaystyle{c_1^\dagger c_1}} & = & \left\vert 
{\displaystyle{\frac{\phi(N_1-1)}{\phi(N_1)}}} \right \vert^2 \\
{\displaystyle{c_2^\dagger c_2}} & = & |q|^{2N_1} \left\vert 
{\displaystyle{\frac{\phi(N_2-1)}{\phi(N_2)}}} \right \vert^2 \\
\vdots & & \end{array}\right\} 
\eqno(4.37)
$$

\noindent By choosing $\phi(N_j)$ one then gets the desired
explicit expressions for N$_j$.  However, as already emphasized, such
expressions are not of much use; the universal representation of N$_j$
given in eq.(4.34) is sufficient for all purposes.

The formalism of Fock space developed here has sufficient flexibility so
as to allow many strightforward extensions or generalizations. For
instance, the relationship between permuted states defined by eqs(4.1)
or (4.2) can be generalized further. Instead of eq(4.3), we may take 
$$
c_j c_i - q^\ast_{ij} c_i c_j \ = \ 0 \quad {\rm for} \quad i > j\,,
\eqno(4.38)
$$

\noindent where $q_{ij}$ are complex parameters. This may be called {\it
multiparameter $q$-statistics}. The algebras corresponding to such
extensions as well as further generalizations can all be constructed
essentially by the same procedure as given in this section. Details will
be presented elsewhere. Suffices it to say that, once the relationship
among the permuted states is specified through the $cc$ relation, the
rest of the story follows. 

Special cases of such algebras have been reported in the literature on
quantum groups [21-33]. We have already referred to the generality of our
approach as compared to quantum groups. As further points of comparison
between the two approaches we must mention the following. From the point
of view of quantum groups, the whole set of relations among $c$ and
$c^\dagger$ are taken for granted. In contrast, our analysis based on the
underlying Fock space reveals the $cc$ relation as the key to the whole
algebraic structure, although the normalization function $\phi(n)$ also
plays a role in determining the actual representation of the operators.
Hence, depending on the mode of expressing the $cc$ relation and the
choice of $\phi(n)$, one can generate any number of algebras of $c$ and
$c^\dagger$. Thus, our approach helps to demystify the quantum - group
related algebras by reducing them to their basic essentials which are
identified to be simple properties of state vectors in the Fock space.
Reversing the procedure, one can probably reconstruct the whole edifice
of the quantum group itself starting from the more elementary notions
relating to states in the Fock space. This of course lies outside the
purview of the present paper.

\vspace{.5cm}

\noindent {\bf 4.2 Fock space of frozen order and null statistics}

In subsection 4.1 we had taken any two states obtained by
permutation to be related to each other. We now consider a limiting
situation in which the particles are frozen in a particular order, with
no permutation allowed. This is the {\it Fock space of frozen
states } and the asosciated statistics is {\it the statistics of frozen
states}.  Whereas in infinite statistics each permutation leads to a new
state, in the statistics of frozen states no permutation is allowed. To
emphasize this contrast, the latter may be called {\it null statistics}.
Although this may be obtained as the limit $q \rightarrow 0$ of
the $q$-statistics, more properly, it must be regarded as an independent
statistics. For, whereas the statistics for two (non-zero) values of $q$
are related to each other through eqs(4.32) or (4.33), the statistics
for $q=0$ cannot be related to statistics for $q\ne 0$. Hence, we shall
construct the algebra of $c$ and $c^\dagger$ for this system of frozen
states directly from the definition. Again, depending on whether the
particles obey exclusion principle or not, we have two different
versions of the system which we shall call the fermionic and the bosonic
versions respectively.

Let us first consider the bosonic version. Referring to eq(4.1) the
Fock space of frozen states is obtained by taking $|n_1 n_2 \ldots ; 1>$
as the allowed state and requiring
$$
|n_1 n_2 \ldots ; \mu> \ = \ 0 \quad {\rm for} \quad \mu \ne 1
\eqno(4.39)
$$

\noindent Or, equivalently
$$
| \ldots 1_j 1_i \ldots > \ = \ 0 \quad {\rm for} \quad i < j
\eqno(4.40)
$$

\noindent and hence 
$$
c_i c_j \ = \ 0 \quad {\rm for} \quad i < j \eqno(4.41)
$$

Assuming factorization of the norm as in eq(4.9), we can represent the
creation operator in the form :
$$
c^\dagger_j \ = \ \sum_{n_j, n_{j+1} \ldots}
\frac{\phi(n_j)}{\phi(n_j+1)} \vert \vert 0_1, \ldots  0_{j-1}, 
(n_j +1), n_{j+1} \ldots \gg \ll 0_1, \ldots 0_{j-1}, n_j, n_{j+1} 
\ldots \vert \vert \eqno(4.42)
$$

\noindent The zeroes in the state vectors arise from the fact that
$c^\dagger_j$ can create a quantum with index $j$ only if indices $i<j$
are unoccupied. From eq(4.41) and the orthogonality of the states, we
get
$$
c_k c^\dagger_j \ = \ 0 \quad {\rm for} \quad k \ne j \eqno(4.43)
$$

\noindent The $c_jc^\dagger_j$ algebra will depend on the choice of
$\phi(n_j)/\phi(n_j+1)$. We shall assume $\phi(n_j) = \phi(n_j+1)$ for
simplicity. Then, by using the completeness relation for the states, one
can verify
$$
c_j c^\dagger_j \ = \ 1 - \sum_{k < j} c^\dagger_k c_k \eqno(4.44)
$$

\noindent Eqs (4.41), (4.43) and (4.44) together define the algebra for
the bosonic version of the statistics of frozen order. 

For the fermionic version, eq(4.41) is replaced by
$$
c_i c_j \ = \ 0 \quad {\rm for} \quad i \le j \eqno(4.45) 
$$

\noindent and one again gets
$$
c_k c^\dagger_j \ = \ 0 \quad {\rm for} \quad k \ne j \eqno(4.46)
$$

\noindent but, instead of eq (4.44), one finds
$$
c_j c^\dagger_j \ = \ 1 - \sum_{k \le j} c^\dagger_k c_k
\eqno(4.47)
$$

\noindent Thus, the fermionic version of the statistics of frozen states
is described by the algebra of eqs(4.45)-(4.47). It is interesting to
note that the replacement of the sign $<$ in the bosonic algebra by the
sign $\le$ yields the fermionic algebra.

Finally we note that the expression for the number operators in terms of
$c$ and $c^\dagger$ for the above algebra of null statistics can be
shown to be essentially the same as that for the standard representation
of infinite statistics (eq 3.9a), but because of eqs (4.41) or (4.45),
it can be simplified to read as
$$
N_k = \sum_{n_1,n_2\dots n_k} c_1^{\dagger n_1}
c_2^{\dagger n_2} \ldots c_k^{\dagger n_k} c_k^{\dagger} c_k
c_k^{n_k} \ldots c_2^{n_2} c_1^{n_1}
\eqno(4.48)
$$

\noindent where the sum over $n_1,n_2 \ldots n_k$ is unrestricted $(\ge
0)$ for the bosonic version, but is restricted for the fermionic
version : $n_1,n_2 \ldots n_{k-1}$ go over 0 and 1 while $n_k = 0$.

\noindent {\bf 4.3 Thermodynamic aspects}

For the bosonic and fermionic Fock spaces, eq.(3.40) for the partition
function becomes
$$
Z \ = \ \sum_{n_1,n_2\ldots} \ 
e^{\displaystyle{{-\beta \sum_i \epsilon_i n_i}}} \eqno(4.49)
$$

\noindent where $n_i \ = \ 0,1,2,\ldots \infty$ for the bosonic space
and $n_i = 0,1$ for the fermionic space. This is a well-known result for
the canonical bose and fermi statistics, but we emphasise the fact that
{\it the partition function and the resulting thermodynamics is the same
for all the various $q$-statistics and $q$-deformed algebras living in
the same bosonic Fock space and similarly for the fermionic Fock space}.
This point has been missed in much of the current literature and so
considerable confusion has been created. Part of this confusion is due
to the choice of the inappropriate Hamiltonian (eq.(3.42)) and the wrong
emphasis placed on some particular algebra of $c$ and $c^\dagger$.

We further note that the above $Z$ in eq(4.49) is valid even for the
Fock space of frozen order and null statistics since $d = 1$ in this
case too.

To sum up, equilibrium thermodynamics of a system of free particles is
determined entirely by the dimension $d$ and the spectrum of allowed
occupation numbers which together characterize the Fock space and does
not depend on the permutation properties of the multiparticle states or
the algebras of $c$ and $c^\dagger$.

\newpage
 

\baselineskip=24pt
\setcounter{page}{51}

\noindent {\bf {5. Derivation of cc relations from $cc^\dagger$
algebras}}

In the last section we showed how to construct the bilinear algebra of c
and c$^{\dagger}$ starting from states related by eq.(4.1).  In other
words, cc$^{\dagger}$ algebra has been derived from the cc algebra given
by eq.(4.3).  The converse is also true ; the cc algebra can be derived
from the cc$^{\dagger}$ algebra, as will be shown in this section.  
Thus, within the
framework of Fock space it is unnecessary to give both cc and
cc$^{\dagger}$ relations.  Either the $cc$ relation or the $cc^\dagger$
relation can be used to define the Fock space and the other can be
derived. But there are some caveats:

\noindent (i) Although the $cc$ relation does define the Fock space, the
$cc^\dagger$ algebra does not follow uniquely ; as already pointed out,
the operators $c,c^\dagger$ and their algebras depend on the choice of
$M$. 

\noindent (ii) In order to define the Fock space completely, the
$cc^\dagger$ algebra must fulfill certain conditions. The $cc^\dagger$
algebra must be such that an arbitrary inner product $<0|c_ic_j \ldots
c^\dagger_g c^\dagger_h |0>$ or infact the vacuum matrix element of any
polynomial in the $c$'s and $c^{\dagger}$ 's arbitrarily ordered, can be
calculated using the $cc^\dagger$ algebra and the definition of the
vacuum state $|0>$ :
$$
c_k |0> \ = \ 0 \quad {\rm for\ all } \quad k\,. \eqno(5.1)
$$

\noindent A general form of the $cc^\dagger$ algebra that satisfies this
requirement is :
$$
c_i c^\dagger_j \ = \ A_{ij} + \sum_{k,m} B_{ijkm} c^\dagger_k c_m
\eqno(5.2)
$$

\noindent where $A_{ij}$ and $B_{ijkm}$ can be constants, but more
generally they can be  functions of the number operators.
All the $cc^\dagger$ algebras considered in this
paper are of this form.

There exists an elegant method [19] to derive cc relations from 
the $cc^{\dagger}$ algebras. Restricting ourselves to relations quadratic in
$c$,  we define
$$
Q_{ij} \, = \, c_i c_j - q^{'} c_j c_i    \eqno(5.3)
$$

\noindent where q$^{'}$ may be an arbitrary complex parameter. Suppose
that, by using the given cc$^{\dagger}$ algebra, we are able to prove
$$
Q_{ij} c^{\dagger}_k \,=\, \sum_{i'j'k'} \, F_{ijk; i'j'k'} \,
c^{\dagger}_{k'} \, Q_{i'j'}    \eqno(5.4)
$$

\noindent for all i,j and k where F$_{ijk;i'j'k'}$ may be a c-number or
operator. Then, by applying this equation successively, we get
$$
Q_{ij} c^{\dagger}_k c^{\dagger}_m \ldots \,=\, \sum_{i'j'k'} \,
\sum_{i''j''m'} \, F_{ijk;i'j'k'} \, F_{i'j'm; i''j''m'} \, c^{\dagger}_{k'}
\, c^{\dagger}_{m'}\ldots Q_{i''j''}    \eqno(5.5)
$$

\noindent Allowing both sides of this equation to act on $\vert 0 >$, the
right side vanishes and so we
see that $Q_{ij}$ acting on any Fock state $c^{\dagger}_k c^{\dagger}_m
\ldots \vert 0 >$ gives zero.  Hence we may write the operator identity
:
$$
Q_{ij} \,=\, 0   \eqno(5.6)
$$
\noindent which is the cc relation sought after.  It may also be pointed
out that often one finds 
$$
F_{ijk;i'j'k'} \, =\, f_{ijk} \delta_{ii'} \, \delta_{jj'} \,
\delta_{kk'}      \eqno(5.7)
$$
\noindent so that eq.(5.4) is simplified to 
$$
Q_{ij} \, c^{\dagger}_k \,=\, f_{ij k} \, c^{\dagger}_k Q_{ij}
\eqno(5.8)
$$

\noindent Thus, the form-invariance of $Q_{ij}$ on being pushed to the 
right of $c^{\dagger}_k$, as explicitly given in eqs.(5.4) or (5.8) is the
necessary  and sufficient condition for the existence of the $cc$
relation.  We may now
apply this method on various $cc^\dagger$ algebras discussed in the
previous sections.

\noindent {\bf{5.1 $q$-mutator algebra with real $q$}}:

The $cc^\dagger$ algebra is
$$
c_i c^{\dagger}_j - q c^{\dagger}_j c_i \, = \, \delta_{ij} \quad ,
\quad \forall i,j.     \eqno(5.9)
$$

\noindent We define
$$
Q_{ij} \,=\, c_i c_j - q' c_j c_i \quad , \quad  \forall  i,j \eqno(5.10)
$$

\noindent From eq.(5.9),
$$
Q_{ij} c^{\dagger}_k \,=\, (1-qq') c_i \delta_{kj} + (q-q') c_j
\delta_{ki} + q^2 c^{\dagger}_k Q_{ij}     \eqno(5.11)
$$
\noindent The form-invariance of $Q_{ij}$ requires
$$
q' = q = \pm 1.       \eqno(5.12)
$$

\noindent So, we get the cc relations 
$$
c_i c_j \pm c_j c_i = 0        \eqno(5.13)
$$
\noindent corresponding to Bose and Fermi statistics ${\rm
at} \, q = \pm 1$.  For $-1 < q < 1$, there are no $cc$ relations and
we have infinite statistics. The inner product matrix $M$ which remains
positive definite for $-1 < q < 1$, develops zero eigenvalues at $q=\pm
1$. (see eq.(3.13)) corresponding to the null states such as
$(c^\dagger_i c^\dagger_j \pm c^\dagger_j c^\dagger_i)|0>$ arising from
eq.(5.13). 

\noindent {\bf{5.2 $q$-mutator algebra with complex $q$}}

The $cc^\dagger$ relations are 
$$
c_i c^{\dagger}_j - q c^{\dagger}_j c_i  =  0   \quad {\rm for} \, i <
j   \eqno(5.14)
$$
$$
c_i c^{\dagger}_i - p c^{\dagger}_i c_i  =  1    \eqno(5.15)
$$

\noindent We define
$$
Q_{ij} \,  =  \, c_i c_j - q^{'} c_jc_i \quad {\rm for} \quad i < j 
\eqno(5.16)
$$
$$
\,  =  \, c_i c_i \quad {\rm for} \quad i = j    \eqno(5.17)
$$

\noindent Using eqs.(5.14) and (5.15), we find
$$
\left.  \begin{array}{lcl}
Q_{ij} c^{\dagger}_k & = & q^2 c^{\dagger}_k Q_{ij} \, \quad for \,
i \le j < k \\
&\\
& = & q^{\ast^2} c^{\dagger}_k Q_{ij} \, \quad for \,  k < i \le j \\
&\\
& = & \vert q \vert^2 c^{\dagger}_k Q_{ij} \, \quad for \,  i <  k < j \\
&\\
& = & (q^{\ast} - q^{'}) c_j + pq^{\ast} c^{\dagger}_i Q_{ij} \, \quad for 
\, k = i < j \\
&\\
& = & (1-qq^{'}) c_i + pq c^{\dagger}_j Q_{ij} \, \quad for \,  i < j = k \\
&\\
& = & (1+p) c_i + p^2 c^{\dagger}_i Q_{ii} \, \quad for \,  i = j = k.
\end{array}    \right\}      \eqno(5.18)
$$

\noindent The form-invariance of $Q_{ij}$ for $i < j$ requires
$$
q^{'} \, = \, q^{\ast} = e^{-i\theta}     \eqno(5.19)
$$

\noindent where $\theta$ is real.  We thus get the cc relation
corresponding to fractional statistics at the boundary $\vert q \vert =$
1 of the disc $\vert q \vert < 1$ in the complex plane.

We already know that the disc $|q|<1$ corresponds to the region of
positive-definite inner product matrix $M$ for the complex $q$-mutator
algebra of infinite statistics. At the boundary of this positivity
region, we have fractional statistics living in a reduced Fock space
(see Fig.2). It is the states of vanishing norm of the form
$$
( \ldots c^{\dagger}_i c^{\dagger}_j \ldots \vert 0 > ) \quad  - e^{i
\theta} (\ldots c^{\dagger}_j c^{\dagger}_i \ldots | 0 > ) \quad {\rm
for} \quad i < j \eqno(5.20)
$$

\noindent which contribute to the zeroes of det M$_n$ in eq.(3.34)  at
$\vert q \vert = $ 1. Once we remove these states, we get the reduced
Fock space of positive-definite norm.

The form-invariance of $Q_{ij}$ \, for i = j requires
$$
p = -1      \eqno(5.21)
$$
\noindent with the corresponding cc relation
$$
c_i c_i \, = \, 0    \eqno(5.22)
$$

\noindent This leads to the exclusion principle namely $n_i = 0$ or 1 only ;
equivalently, the norms of states with repeated indices are zero as seen 
by putting $p = -1$ in eq.(3.35).

To sum up, we may note four possibilities all contained in the algebra
of eqs (5.14) and (5.15) :

\noindent (a) Infinite statistics with multiple occupation
$$
\vert q \vert \neq 1 \quad ; \quad p \neq - 1.     \eqno(5.23)
$$

\noindent (b) Infinite statistics with exclusion principle
$$
\vert q \vert \neq 1 \quad ; \quad p = -1     \eqno(5.24)
$$
\noindent (c) Fractional statistics with multiple occupation
$$
q = e^{i\theta} \quad ; \quad p \neq -1     \eqno(5.25)
$$
\noindent (d) Fractional statistics with exclusion principle
$$
q = e^{i \theta} ; p = - 1.   \eqno(5.26)
$$

\noindent {\bf{5.3 $q$-statistics }}

The cc$^{\dagger}$ relations are
$$
c_i c^{\dagger}_j - q c^{\dagger}_j c_i \, = \, 0 \quad {\rm for} \quad
i < j  \eqno(5.27) 
$$
$$
c_j c^{\dagger}_j - p c^{\dagger}_j c_j \, = \, |q|^{2
\sum_{i<j} N_i} f(N_j) \eqno(5.28)
$$
\noindent We define
$$
Q_{ij} \, = \, c_i c_j - q' c_j c_i \quad {\rm for} \quad i > j 
\eqno(5.29)
$$

\noindent Using eqs.(5.27), (5.28) we get

$$
\left.  \begin{array}{lcl}
Q_{ij} c^{\dagger}_k & = & q^2 c^{\dagger}_k Q_{ij}   \quad  
{\rm for}  \quad i <  j < k \\
& \\
& = & q^{\ast^2} c^{\dagger}_k Q_{ij} \,  \quad {\rm for} \quad  k < i < j \\
& \\
& = & \vert q \vert^2 c^{\dagger}_k Q_{ij} \,  \quad  {\rm for} \quad  
i <  k < j \\
& \\
& = & p q^\ast c^{\dagger}_j Q_{ij} + c_i |q|^{2\sum_{m < j}
N_m} f(N_j) (1-q'q^\ast)  \quad {\rm for} \quad i > j = k \\
& \\
& = & pq c^{\dagger}_i Q_{ij} + \vert q\vert^{2\sum_{m < i}N_m} f(N_i) 
c_j(q-q'|q|^2) \quad {\rm for} \quad k = i > j 
\end{array}    \right\}    \eqno(5.30)
$$

\noindent Repeating the argument of form-invariance of $Q_{ij}$, 
we conclude
$$
c_i c_j - q^{\star^{-1}} c_j c_i = 0 \quad {\rm for} \quad i > j   
\eqno(5.31)
$$

\noindent In this proof we have not used any particular form of $f(N_j)$,
but have used only the standard commutation relations among $N_k$ and
$c_i$ (eqs(2.23) and (2.26).

This derivation of the $c_ic_j$ relation for $i \ne j$ is valid for the
bosonic as well as the fermionic fock spaces. However, for the fermionic
space, there exists the additional relation :
$$
c_ic_i \ = \ 0 \eqno(5.32)
$$

To derive this, we now commute $c^\dagger_k$ through $c_ic_i$ 
using the same eqs(5.27 - 5.28), we get
$$ \left. \begin{array}{lcl}
c_ic_ic^\dagger_k & = & q^{\ast 2} c^\dagger_k c_i c_i \quad {\rm for}
\quad i > k \\
& = & q^2 c^\dagger_k c_i c_i \quad {\rm for} \quad k > i \\
\\
& = & p^2 c^\dagger_i c_i c_i + |q|^{2 \sum_{m<i} N_m} c_i \{ pf(N_i-1) +
f(N_i)\} \ {\rm for} \ i=k \end{array} \right\} \eqno(5.33)
$$

\noindent So, for the validity of eq(5.32), we require
$$
c_i\{pf(N_i-1) + f(N_i)\} \ = \ 0 \eqno(5.34)
$$

\noindent Substituting the form of $f(N_i)$ for the fermionic Fock space
(from eq.(4.24))
$$
f(N_i) \ = \ |r|^2 (\delta_{N_i,0} - p \delta _{N_i,1})\,,
\eqno(5.35)
$$

\noindent we see that eq.(5.34) is satisfied, since in the fermionic
Fock space, $\delta_{N_i,2} = 0$ and $c_i\delta_{N_i,0} = 0.$

One may also note that eq.(5.34) is also satisfied for $p=-1$ and
$f(N_i)$ = constant (which may be chosen to be unity). So, for bosonic
Fock space, i.e. $c_ic_i \ne 0$, we must avoid $p=-1$ in eq.(5.28) and
(4.15).

The above derivations of $cc$ relations have used the general
$cc^\dagger$ algebra and hence includes the cases of
quantum-group-covariant algebras, independent deformed oscillators,
commuting fermions, anticommuting bosons etc. Further, it must be noted
that the presence of the term with the factor
 $|q|^{2 \sum_{i<j} N_i}$ in eq.(5.28) is crucial for
the validity of the $cc$ relations ; without this factor, there will be
no $cc$ relation and we will get infinite statistics.

\newpage


\baselineskip=24pt
\setcounter{page}{59}

\noindent {\bf 6. Two-indexed Systems}

\vspace{.5cm}

We have so far considered generalized Fock spaces consisting of states $|n_g,
n_h, \ldots ; \mu >$ where the indices $g,h,\ldots$ may refer either to
a single quantum number, or to a collection of quantum numbers,
specifying the space, spin and other internal degrees of freedom. In the
latter case, it may be supposed that one has mapped a collection of indices 
to a single index.  What we are envisaging in this
section are situations where such mapping is not possible. This can
happen in various ways. To be specific, let us consider oscillators with
a pair of indices, a latin index $(g,h,\ldots)$ and a greek index
$(\alpha, \beta \ldots)$. There exists a class of systems in which the
occupation numbers with a single index $n_g$ or $n_\alpha$ are defined,
but occupation numbers with both the indices $n_{g\alpha}$ are not
defined. Such systems cannot be mapped into single-indexed systems. In
another class of systems, $n_{g\alpha}$ do exist, but the subsidiary
conditions that define the reduced Fock space depend on the two indices
$g$ and $\alpha$ in such a way that prevents mapping of $(g,\alpha)$
into a single index. We consider these two classes of systems in
subsections 6.1 and 6.2 respectively.

It may also be mentioned that we first encountered such systems in the
study of the Hubbard model in the limit of infinite Coulomb repulsion
[43-46]. Although this was our original motivation, this has now opened the
door to a more general framework encompassing novel forms of statistics
and algebras.

\vspace{1cm}

\noindent {\bf 6.1 \ Systems in which $n_{g\alpha}$ do not exist}

We specify the states as $|n_g,n_h\ldots ; n_\alpha, n_\beta \ldots ;
\mu>$, where $n_g, n_h$ $\ldots$ are the numbers of quanta with indices $g,h$
$\ldots$ respectively while $n_\alpha, n_\beta, \ldots$ are the numbers of
quanta with indices $\alpha, \beta\ldots$ respectively and we have the
constraint :
$$
n_g + n_h + \ldots \ = \ n_\alpha + n_\beta + \ldots \eqno(6.1)
$$

\noindent We may regard $n_g$ as the total number of quanta with index
$g$ whatever may be their greek index and similarly for $n_\alpha$.
In such a state, the occupancy number with both indices such as
$n_{k\alpha}$ is not defined. The latin indices and the greek indices
are independently permutated and this leads to a much enlarged space in
each sector. Now $\mu$ goes over $1 \ldots s'$ where
$$
s' \ = \ \frac{(n_g + n_h + \ldots)!}{n_g ! n_h! \ldots } \times 
\frac{(n_\alpha + n_\beta + \ldots)!}{n_\alpha ! n_\beta! \ldots } \eqno(6.2)
$$

\noindent This is to be compared to the size of the space of the states
$|n_{g\alpha}, n_{h\beta} \ldots ; \mu>$ for which the range of $\mu$ is
$$
s \ = \ \frac{(n_{g\alpha} + n_{h\beta} + \ldots)!}{n_{g\alpha} ! 
n_{h\beta }! \ldots }  \eqno(6.3)
$$

\noindent In general, $s'$ is larger than $s$.

Let us consider the two-particle sector as an example. If we specify
occupation numbers with both indices, we have the two states in the
sector $(1_{k\alpha}, 1_{m\beta}) \ : \ | 1_{k\alpha}, 1_{m\beta} ; \mu>$
with $\mu = 1 $ or $2$ which correspond to $|1_{k\alpha}, 1_{m\beta}>$
and $|1_{m\beta}, 1_{k\alpha}>$ and another two states in the sector 
$(1_{m\alpha}, 1_{k\beta}) : | 1_{m\alpha}, 1_{k\beta} ; \mu>$ with $\mu
= 1$ or $2$ corresponding to $|1_{m\alpha}, 1_{k\beta}>$ and
$|1_{k\beta}, 1_{m\alpha}>$. (Here we are not using $k\alpha, m\beta$ etc
in their generic sense ; they are used to denote specific values of the
indices.) On the other hand, with the new type of states with decoupled
indices for which
occupation numbers with both indices do not exist, we have the
two-particle states $|1_k, 1_m ; 1_\alpha, 1_\beta ; \mu>$ and now $\mu$
goes over 1 to 4 corresponding to the four state vectors $|1_{k\alpha},
1_{m\beta}>, \linebreak |1_{m\beta}, 1_{k\alpha}>, |1_{m\alpha}, 1_{k\beta}>,
|1_{k\beta}, 1_{m\alpha}>$. Thus, both the sectors considered earlier
are combined to form a single enlarged sector. Such a regrouping of
sectors with consequent enlargement occurs throughout the Fock space,
in the case of the decoupled indices.

The construction of the orthonormal set as well as the other properties
of the generalized Fock spaces given in Sec.2 goes through for the present
case of decoupled indices also except that the matrices $X,M$ etc will
be of higher dimensions. The creation and destruction operators
$c^\dagger_{k\alpha}, c_{k\alpha}$ also can be constructed in the same
way as in Sec.2. The relevant equations and formulae with the
appropriate changes incorporated are given in Appendix A. They are
self-explanatory.

Just as in the case of single-indexed systems (cf.Sec.2), we can again
have a super Fock space in which all the states connected by independent
permutation of the latin and greek indices are taken to be independent.
We shall call the associated statistics as ``doubly-infinite''
statistics since it is infinite statistics in latin and greek indices
separately. By imposing relations among the permutated states, one can
get many kinds of reduced Fock spaces. Because of the larger number of
available states in each sector, many new types of statistics become
possible. These can be discussed and the associated algebras can be
constructed by the same procedure as in Sec.3, 4 and 5. However, we
shall be brief and restrict ourselves to presenting some of the
important results only. Some of these algebras and the new kinds of
statistics implied by them have been discussed by us in greater detail
in the earlier papers [18,19,46].

\noindent {\bf Doubly-infinite statistics}

Consider the $cc^\dagger$ algebra described by
$$
c_{k\alpha} c^\dagger_{m\beta} - q \delta_{\alpha \beta} \sum_\gamma
c^\dagger_{m\gamma} c_{k\gamma} \ = \ \delta_{km} \delta_{\alpha\beta} 
\eqno(6.4)
$$

\noindent where $q$ is a real parameter lying in the range
$$
-1 < q < 1\,. \eqno(6.5)
$$

\noindent It is the second term on the left of eq(6.4) in which $\alpha$
and $\beta$ have been dissociated from $k$ and $m$ respectively that
leads to the decoupling of the latin and greek indices and prevents the
mapping of lattin and greek indices to a single index. One can show
[19] that the inner-product matrices following from this algebra are
all positive definite for $-1<q<1$. Further, one can show using 
arguments similar to those in Sec.5 that there is no $cc$ relation for
this algebra in the same parameter range and so all the states connected
by {\it independent} permutation of the latin and greek indices are
independent. The underlying Fock space is the full super Fock space of
dimension given by eq.(6.2) and we have infinite statistics in latin and
greek indices separately, or doubly-infinite statistics. 

For $q=0$ in eq(6.4), the two indices can be mapped into a single index
and the algebra reduces to the standard representation of single-indexed
infinite statistics described in Sec.3.1.

The algebra of eq.(6.4) is covariant under the unitary
transformations on the latin indices :
$$
d_{k\alpha} = \sum_m U_{km} c_{m\alpha} \ ; \ U^\dagger U \ = \
UU^\dagger \ = \ 1 \eqno(6.6)
$$

\noindent as well as under separate unitary transformations on the greek
indices :
$$
e_{k\alpha} \ = \ \sum_{\lambda} V_{\alpha\lambda} c_{k\lambda} \ ; \
V^\dagger V \ = \ VV^{\dagger} = 1. \eqno(6.7)
$$

\noindent But the algebra is not covariant under the enlarged unitary
transformations involving both the latin and greek indices (for $q \ne
0$). A special case of the unitary transformations of eqs(6.6) and (6.7)
is the phase transformation. Eq.(6.4) is covariant under either of the
following phase transformations :
$$
d_{k\alpha} \ = \ e^{i \phi_k} c_{k\alpha} \eqno(6.8)
$$
$$
e_{k\alpha} \ = \ e^{i \phi_\alpha} c_{k\alpha} \eqno(6.9)
$$

\noindent As a consequence, the number operators $N_k$ and $N_\alpha$
exist. However, eq(6.4) is not covariant under the transformation :
$$
f_{k\alpha} \ = \ e^{i \phi_{k\alpha}} c_{k\alpha} \eqno(6.10)
$$

\noindent and correspondingly, $N_{k\alpha}$  does not exist.

We can make $q$ in (6.4) complex provided we order the latin indices and
thus we get an algebra which is the analogue of the $q$-mutator algebra
with complex $q$ for the single-indexed systems (Sec.3) :
$$
c_{i\alpha} c^\dagger_{j\beta} - q \delta_{\alpha\beta} \sum_\gamma
c^\dagger_{j\gamma} c_{i\gamma} \ = 0 \quad {\rm for} \quad i < j 
\eqno(6.11)
$$
$$
c_{j\alpha} c^\dagger_{j\beta} - p \delta_{\alpha\beta} \sum_\gamma
c^\dagger_{j\gamma} c_{j\gamma} \ = \ \delta_{\alpha\beta} \eqno(6.12)
$$

\noindent where $q$ is complex, but $p$ is real. This again describes 
the same doubly-infinite statistics ;
only the representation and algebra are different. This algebra is no
longer covariant under the unitary transformations of eqs (6.6) and
(6.7), but is still covariant under the phase transformations of eqs
(6.8) and (6.9). Positivity of the inner-product matrices $M$ requires
$|q|<1$. As for $p$, similar statements as in Sec.3 can be made.

Coming back to eq(6.4), at the boundary of the range of $q$ given in
eq(6.5), namely at $q = \pm 1$, we get two new forms of statistics
called orthobose and orthofermi statistics [18,19,46] which reside in 
reduced Fock spaces defined by $cc$ relations. The algebras for these 
are given below :

\noindent {\bf Orthobose Statistics} 
$$
c_{k\alpha} c^\dagger_{m\beta} - \delta_{\alpha\beta} \sum_\gamma
c^\dagger_{m\gamma} c_{k\gamma} \ = \ \delta_{km} \delta_{\alpha\beta} \eqno(6.13)
$$
$$
c_{k\alpha} c_{m\beta} - c_{m\alpha} c_{k\beta} \ = \ 0\,.
\eqno(6.14)
$$

\noindent {\bf Orthofermi Statistics}
$$
c_{k\alpha} c^\dagger_{m\beta} + \delta_{\alpha\beta} \sum_\gamma
c^\dagger_{m\gamma} c_{k\gamma} \ = \ \delta_{km} 
\delta_{\alpha\beta} \eqno(6.15)
$$
$$
c_{k\alpha} c_{m\beta} + c_{m\alpha} c_{k\beta} \ = \ 0\,.
\eqno(6.16)
$$

The inner product matrices $M$ for the reduced Fock spaces generated by
these algebras (eqs(6.13) - (6.16)) can be shown to be positive. In both
these statistics, any two states obtained by permuting the greek indices
are independent. On the other hand, states obtained by permuting the
latin indices are either equal to each other, or related by $(-1)^J$
where $J$ is the number of inversions in the permutation of the latin
indices, for the orthobose or orthofermi statistics respectively. Thus,
we have infinite statistics in the greek indices and bose or fermi
statistics in the latin indices. For orthofermi statistics, we have the
further condition :
$$
n_k \ = \ 0 \ {\rm or} \ 1 \ {\rm only}\,. \eqno(6.17)
$$

\noindent It must be noted that the exclusion implied by eq(6.17) is
stronger than the usual Pauli exclusion principle ; eq(6.17) requires
that there cannot be more than one particle with index $k$ whatever may
be its greek index.

We have constructed a local relativistic quantum field theory [47] based
on orthostatistics. It may be worth mentioning that although
orthostatistics does involve infinite statistics, the problems faced by
infinite statistics are avoided by associating the latin index
(pertaining to fermi or bose statistics) with the conventional degrees
of freedom such as momentum and spin and assigning the greek index (of
infinite statistics) to a new degree of freedom.

Orthostatistics can be generalised to $q$-orthostatistics. We keep the
infinite statistics in the greek indices but have $q$-bose or $q$-fermi
statistics in the latin indices. More precisely, we must regard these as 
$q$-orthostatistics lying in orthobosonic and orthofermionic Fock
spaces. We can call them $q$-orthobose or $q$-orthofermi statistics.

\noindent {\bf $q$ - Orthobose Statistics}
$$
c_{i\alpha} c^\dagger_{j\beta} + q \delta_{\alpha\beta} \sum_\gamma
c^\dagger_{j\gamma} c_{i\gamma} \ = \ 0 \quad {\rm for} \quad i < j \eqno(6.18)
$$
$$
c_{j\alpha} c^\dagger_{j\beta} - |q|^2 \delta_{\alpha\beta} \sum_\gamma
c^\dagger_{j\gamma} c_{j\gamma} \ = \ \delta_{\alpha\beta} + \delta_{\alpha\beta}
(|q|^2-1) \sum_{k<j} \sum_\gamma c^\dagger_{k\gamma} c_{k\gamma} \eqno(6.19)
$$
$$
c_{i\alpha} c_{j\beta} + q^\star c_{j\alpha } c_{i\beta} \ = \ 0 \quad
{\rm for } \quad i < j \eqno(6.20)
$$

\noindent {\bf $q$-Orthofermi Statistics}
$$
c_{i\alpha} c^\dagger_{j\beta} + q \delta_{\alpha\beta} \sum_\gamma
c^\dagger_{j\gamma} c_{i\gamma} \ = \ 0 \quad {\rm for} \quad i < j \eqno(6.21)
$$
$$
c_{j\alpha} c^\dagger_{j\beta} + \delta_{\alpha\beta} \sum_\gamma
c^\dagger_{j\gamma} c_{j\gamma} \ = \ \delta_{\alpha\beta} + 
\delta_{\alpha\beta}
(|q|^2-1) \sum_{k<j} \sum_\gamma c^\dagger_{k\gamma} c_{k\gamma} \eqno(6.22)
$$
$$
c_{i\alpha} c_{j\beta} + q^\star c_{j\alpha} c_{i\beta} \ = \ 0 \quad
{\rm for } \quad i < j \eqno(6.23)
$$
$$
c_{j\alpha} c_{j\beta} \ = \ 0 \eqno(6.24)
$$

In eqs(6.18) - (6.24), $q$ is an arbitrary complex parameter. For
$q=e^{i\theta}$, and $q=0$ we shall have fractional statistics and
statistics of frozen order respectively, but in the latin indices only.
Of course one can construct many other algebras corresponding to the
same statistics, just as in the case of the single-indexed systems.

Further, these equations are analogous to eqs(4.3), (4.12),(4.23) and
(4.28) which are covariant under quantum groups. So, another direction
is indicated here for the further generalization of
quantum-group-theoretic structures to two-indexed systems.

\vspace{1cm}

\noindent {\bf 6.2 Systems in which $n_{g\alpha}$ exist}

We shall now consider a different kind of double-indexed systems. Here,
the occupation numbers $n_{g\alpha}$ exist, nevertheless mapping to
single-indexed systems is not possible because of the subsidiary
conditions that define the reduced Fock space. We give below
three examples of such double indexed systems, described by the algebras
:
\begin{description}
\item[(a)] 
$$
c_{k\alpha} c^\dagger_{m\beta} + (1-\delta_{km})
c^\dagger_{m\beta} c_{k\alpha} \ = \ \delta_{km} \delta_{\alpha\beta}
\left( 1 - \sum_\gamma c^\dagger_{k\gamma} c_{k\gamma}\right) \eqno(6.25)
$$
\item[] 
$$
c_{k\alpha} c_{m\beta} + (1-\delta_{km}) c_{m\beta} c_{k\alpha}
\ = \ 0\,. \eqno(6.26)
$$
\item[(b)] 
$$
c_{k\alpha} c^\dagger_{m\beta} - (1-\delta_{km})
c^\dagger_{m\beta} c_{k\alpha} \ = \ \delta_{km} \delta_{\alpha\beta}
\left( 1 + \sum_\gamma c^\dagger_{k\gamma} c_{k\gamma}\right) \eqno(6.27)
$$
\item[]  
$$
(1-\delta_{km}) (c_{k\alpha} c_{m\beta} - c_{m\beta} c_{k\alpha})
\ = \ 0 \,. \eqno(6.28)
$$
\item[(c)] 
$$
c_{k\alpha} c^\dagger_{m\beta} - (1-\delta_{km})
c^\dagger_{m\beta} c_{k\alpha} \ = \ \delta_{km} \delta_{\alpha\beta}
\left( 1 - \sum_\gamma c^\dagger_{k\gamma} c_{k\gamma}\right) \eqno(6.29)
$$
\item[] 
$$
c_{k\alpha} c_{m\beta} - (1-\delta_{km}) c_{m\beta} c_{k\alpha}
\ = \ 0 \,. \eqno(6.30)
$$
\end{description} 

Each of these algebras is covariant under the unitary transformation on
the greek indices defined by eq.(6.7), but not covariant under the
unitary transformation on the latin indices defined by eq.(6.6).
However, all of them are covariant under not only the phase
transformations in eqs(6.8) and (6.9) but also the $k\alpha$-dependent
phase transformation of eq.(6.10). hence all the occupation numbers
$n_k, n_\alpha$ as well as $n_{k\alpha}$ exist.

Algebra (a) leads to states which are antisymmetric for simultaneous
interchange of latin and greek indices as in Fermi-Dirac statistics, but
the usual Pauli exclusion principle is replaced by the stronger or more
exclusive exclusion principle, as in orthofermi statistics (see eq.6.17)
:
$$
n_k  \ = \ 0 \quad {\rm or} \quad 1 \quad  {\rm only} \,, \eqno(6.31)
$$

\noindent where
$$
n_k \ \equiv \ \sum_\alpha n_{k\alpha} \eqno(6.32)
$$

\noindent These are precisely the states that are allowed in the Hubbard
model of strongly correlated electrons in the limit of infinite
intrasite Coulomb repulsion if we interpret the latin index as the site
and the greek index as the spin. Hence, this algebra and the statistics can
be called Hubbard algebra and Hubbard statistics respectively. For more
details, the reader is referred to [46]. Since $n_{k\alpha}$ exists, the
states of the system can be characterized as $|n_{k\alpha}, n_{m\beta}
\ldots ; \mu>$. However, because of the above constraint (eq(6.31))
which can be rewritten in the form :
$$
\sum_\alpha n_{k\alpha} \ = \ 0 \quad {\rm or} \quad 1 \quad {\rm
only} \eqno(6.33)
$$

\noindent this sytem cannot be mapped into a single-indexed system.

Algebra (b) can be regarded as the bosonic ``counterpoint'' of algebra
(a). States are symmetric under simultanueous exchange of latin and
greek indices for quanta with  different latin indices $(k \ne m)$. But
for quanta with identical latin indices $(k = m)$, there is no
restriction on the symmetry with respect to the greek indices. In other
words, there is infinite statistics in greek indices if the
corresponding latin indices are identical. Whereas algebra (a) leads to
more exclusive states than allowed by Pauli, algebra (b) leads to ``more
inclusive'' states than allowed by Bose. For this reason we
may call algebra (b) as the ``inclusive counterpoint'' to algebra (a).
Such a restriction on the allowed states of a two-indexed system which
distinguishes $k=m$ from $k\ne m$ cannot be mapped into a condition for
a single-indexed system.

Finally, algebra (c) leads to states that are symmetric for simultaneous
exchange of latin and greek indices, but the stronger exclusion
principle of eq (6.33) is also valid as in algebra (a). Hence, algebras
(a) and (c) represent two forms of statistics which may be called
antisymmetric and symmetric Hubbard statistics respectively 
both residing within the same reduced Fock space. 
On the other hand, algebra (b) and its statistics lie in a
different reduced Fock space which is a Fock space with the new
``inclusion'' principle.

A compilation of the algebras and statistics for two-indexed systems is
given in Table IV. The $cc$ relations are not included since they can be
shown to follow from the $cc^\dagger$ algebra, whenever they exist.

\newpage


\baselineskip=24pt
\setcounter{page}{71}
\noindent{\bf 7. Summary and Discussion}

We have formulated a theory of generalized Fock spaces which is
sufficiently general so as to encompass the well known Fock spaces and
many newer ones.  We have shown that such a theory can be constructed
without introducing creation and annihilation operators.  The only
requirements for constructing a generalized Fock space are to specify the
set of allowed states, and to make it an inner product space. By
freeing the notion of the underlying state space from c and
c$^{\dagger}$, we are able to define different forms of quantum
statistics in a representation independent manner.  Subsequently, one
can construct c and c$^{\dagger}$ and their algebras in any desired
representation.

Our general formalism not only unifies the various forms of statistics
and algebras proposed so far but also allows one to construct many new
forms of quantum statistics as well as algebras of c and c$^{\dagger}$
in a systematic manner.  Some of these are the following :

\noindent (a) Many new algebras for infinite statistics \\
\noindent (b) Complex q-statistics and a number of cc$^{\dagger}$
algebras representing them \\
\noindent (c) A consistent algebra of c and c$^{\dagger}$ for
``fractional" statistics \\
\noindent (d) Null statistics or statistics of frozen order \\
\noindent (e) ``Doubly-infinite" statistics and its representations \\
\noindent (f) q-orthobose and q-orthofermi statistics \\
\noindent (g) A statistics for two-indexed systems with a new
``inclusion principle''. \\
\noindent (h) A symmetric version of Hubbard statistics.

Our primary concept is that of generalized Fock space, of which many
categories have been introduced in this paper.  Next comes the notion
of statistics which is defined by the type of symmetry or relationship
among the state vectors residing in the particular type of Fock space.
In a given Fock space, more than one type of symmetry can be postulated,
the prime example of this being the symmetry, antisymmetry or q-symmetry
in the bosonic and fermionic Fock spaces. For a given statistics,
there can exist different representations of c and c$^{\dagger}$,
leading to different $cc^{\dagger}$ relations.  To summarize, a
particular Fock space can admit different statistics, and a particular
statistics can be represented by more than one $cc^{\dagger}$
algebra.  But {\it {the important point is that various
statistics and algebras residing in a given Fock space are all
inter-related}}. These interconnections are given by generalized
versions of the well-known Jordan-Wigner-Klein transformations. 
No such interconnections exist among statistics and
algebras belonging to distinct Fock spaces. We must further add that
equilibrium thermodynamics of a system of free particles is the same for
all the different statistics and algebras within the same Fock space.

For the sake of clarity, the above-described logical order of concepts
as well as their logical interconnections are presented in the form of
flow charts or block diagrams in Figs.3 and 4.  The single-indexed
systems are considered in Fig.3.  The Fock spaces of higher dimension
are shown to the right of those of lower dimension .  The Fock space of
frozen order as well as the bosonic and fermionic Fock spaces have the
lowest dimension $d = 1$ in any sector $\{n_g, n_h \ldots\}$.  Next come the
parafermionic and parabosonic Fock spaces which have d $>$ 1.  At the
extreme right, we have the super Fock space which has the largest
dimension $d = s$ in each sector with s given by eq.(2.6).  Null
statistics and infinite statistics can be regarded as the opposite
limiting cases of generalized statistics and hence these two forms of
statistics along with their Fock spaces occupy the opposite ends of the
diagram.  Although not shown separately in Fig.3 because of lack of
space, the bosonic and fermionic Fock spaces are distinct and each must
be separately associated with the complete set of statistics and
algebras shown.  Same is true of the parabosonic and parafermionic Fock
spaces.  Further, there are two Fock spaces of frozen order, the bosonic
and fermionic type.  And finally, there exists another super Fock space 
with exclusion principle, which is not shown separately.

Within the parafermionic and parabosonic Fock spaces many
``deformations" of parastatistics and many other representations and
algebras apart from Green's trilinear algebra [35] are possible.
These are indicated by the hanging arrows in Fig.3.  Further, as shown
by the dotted lines, there is enough room for many new varieties of Fock
spaces and associated statistics and algebras.  These possibilities may
be pursued in the future.

Coming to Fig.4 depicting the systems with two indices, here again Fock
spaces of higher dimension generally lie to the right.  Although shown
together, the orthobosonic and orthofermionic Fock spaces must be  regarded
distinct.  Here, one can envisage a richer harvest of new Fock spaces,
statistics and algebras because of the two indices and this again is for
the future.

We now conclude with some general comments :

\noindent 1. We must once again repeat and emphasize the point that most of the
$q$-deformations on oscillators discussed in the literature amount to
only a change of variable and hence must be regarded as different
avatars of bosonic or fermionic systems. However, one must clearly
distinguish those deformations such as the $q$-mutator algebra of
Greenberg that require the construction of new types of Fock spaces.
Obviously, Greenberg-type of deformations can never be reduced to change
of variables living within the bosonic or fermionic Fock space. Some
degree of confusion prevails in recent literature since this distinction
is not kept in mind. (See for instance [7,32,41,42,48]).

\vspace{.5cm}

\noindent 2. In Sec.4, we have shown that algebras that are covariant
under quantum groups are only a particular case of the more general
class of algebras that can be derived from the formalism of generalized
Fock spaces.  This formalism is based on linear vector space and linear
operators acting on this space; mathematically, no more sophistication
is required.  And yet it is capable of handling quantum-group related
structures in a self-contained manner. It would seem that the basic
concepts of quantum groups are contained in the theory of generalized
Fock spaces and it must be possible to construct quantum group itself
starting from this theory.

\vspace{.5cm}

\noindent 3. We have already referred to the desirability of covariance under
unitary transformations that mix the indices as a requirement for the
algebras of creation and annihilation operators. We shall call the
algebras that satisfy this requirement as covariant algebras. This
property stems from the superposition principle in quantum mechanics.
Since the indices describe quantum states of a single particle, if we
demand that, for any orthonormal set of quantum states obtained by
superposition of the original set of quantum states, the algebra should
retain the same form, then covariance under unitary transformations
follows. Many of the algebras presented in this paper violate this
requirement.  Nevertheless, these algebras may be useful to
describe specific systems in specific states such as those encountered
in condensed matter physics.

Some of these noncovariant algebras do have other nice physical and
mathematical properties. Under this category, we may include the
algebras that are covariant under quantum groups or the algebras
representing braid group statistics [49,50].

Among the algebras for single-indexed systems that have been 
discussed, Greenberg's $q$-mutator
algebra is the only $q$-deformation that is covariant under unitary
transformations, but then one has to pay the price of the enlarged Fock
space. Every other known $q$-deformation leads to a noncovariant
algebra.

Greenberg's $q$-mutator algebra (including the case $q=0$ which is the
standard representation), the canonical bosonic and fermionic algebras
and Green's trilinear algebras for parabosons and parafermions are the
covariant representatives living respectively in the super Fock space,
bosonic and fermionic Fock spaces and the parabosonic and parafermionic
Fock spaces. All the other algebras living in these three catagories of
Fock spaces, although noncovariant, can be transformed to these
covariant algebras through equations such as eq(4.32). This is not the
case for the algebra of null statisitcs or the algebra of infinite
statistics with Pauli principle living respectively in the Fock space of
frozen order and the super Fock space with Pauli principle. In these Fock
spaces, covariant algebras do not exist.

\vspace{.5cm}

\noindent 4. Quantum mechanics is sometimes viewed as a deformation of
classical mechanics since the commutator bracket of quantum mechanics
can be related to the deformation of the classical Poisson bracket, the
Planck's constant playing the role of the deformation parameter.
Relying on similar reasoning it has been proposed that a deformation of
canonical commutation relations will lead to fundamentally new
mathematical or physical structures [51,52,53].  The analysis presented in
this paper shows that nothing of this sort happens, if viewed within the
framework of Fock space.  The transition from
classical to quantum mechanics requires the replacement of
the notion of the phase space by that of the Hilbert space or Fock
space.  In contrast, we have seen that all the deformations of
commutation relations can be formulated within the framework of Fock
space.  In fact most of the deformed structures proposed in the
literature exist within the time-honoured bosonic and fermionic Fock
spaces only. Even Greenberg's infinite statistics lives within a Fock
space, although an enlarged one. 

\vspace{.5cm}

\noindent 5. While remaining within the framework of quantum mechanics,
the general theory of Fock spaces presented here throws light on the
enlarged framework within which the familiar quantum field theory and
statistical mechanics reside and hence may lead to newer forms of
quantum field theory and statistical mechanics.  This is infact the 
main motivation behind
our work.  Apart from earlier work on parastatistics [37], we may
mention as examples of new forms of quantum field theories, Greenberg's
construction [13,14,54] of a nonrelativistic quantum field theory based 
on infinite statistics and our construction [47] of a local 
relativistic quantum field theory based on orthostatistics.  
Many other forms of quantum field
theories based on the generalized Fock spaces may be possible.  Their
formulation and study is an agenda for the future.

\vspace{.5cm}

\noindent 6. One may not be able to construct local relativistic
quantum field theories corresponding to many of the newer forms of
statistics and algebras, since admissible statistics in relativistic
systems is severely restricted by the axioms of local quantum physics
[55]. However, nonrelativistic quantum field theories based on
such ``inadmissible'' statistics are still possible. Condensed 
matter physics is a rich field where
applications of such theories may be relevant. In fact there is no
reason why the quasiparticles encountered in condensed matter systems
should be bosons or fermions only. We have shown that any of the
generalized Fock spaces provides a perfectly valid quantum-mechanical
framework for many-particle systems. Hence, quasiparticles living in a
generalized Fock space offer an important field of study.

\newpage

\setcounter{page}{79}
\baselineskip=24pt

{\centerline{\bf Appendix A : Generalised Fock Spaces for Two-indexed
Systems}}

\bigskip

Here we consider only those two-indexed systems in which $n_{g\alpha}$ do
not exist. See Sec.6.1.

\medskip

\noindent{\underline {The state vectors, inner products and orthonormal
sets}}
$$
\langle n^{'}_g, n^{'}_h \ldots; n^{'}_{\alpha}, n^{'}_{\beta} \ldots ;
\mu \vert n_g, n_h \ldots ; n_{\alpha}, n_{\beta} \ldots; \nu \rangle
$$
$$
\quad = \quad \delta_{n^{'}_g n_g}  
\delta_{n'_h n_h} \, \ldots  \delta_{n^{'}_{\alpha} n_{\alpha}}  
\delta_{n^{'}_{\beta} n_{\beta}} \ldots M_{\mu \nu}    \eqno(A.1)
$$
$$
\parallel n_g, n_h \ldots; n_{\alpha}, n_{\beta} \ldots ; \mu \gg = 
\sum_{\nu} \, X_{\nu \mu} \, 
\vert n_g, n_h \ldots; n_{\alpha}, n_{\beta} \ldots ; \nu \rangle
\eqno(A.2)
$$
$$ 
\ll  n^{'}_g, n^{'}_h \ldots; n^{'}_{\alpha}, n^{'}_{\beta} \ldots ;
\mu \parallel  n_g, n_h \ldots ; n_{\alpha}, n_{\beta} \ldots; \nu \gg
$$
$$
\quad = \quad \delta_{n^{'}_g n_g}  
\delta_{n^{'}_h  n_h} \, \ldots  \delta_{n^{'}_{\alpha} n_{\alpha}}  
\delta_{n^{'}_{\beta} n_{\beta}} \ldots \delta_{\mu \nu}    \eqno(A.3)
$$
$$
M^{-1} \, = \, XX^{\dagger}     \eqno(A.4)
$$
$$
I \, = \, \sum_{{\stackrel{n_g,n_h ..}{n_{\alpha}, n_{\beta}..}}}
\, \sum_{\mu} \parallel n_g, n_h ..; n_{\alpha}, n_{\beta} .. ;
\mu \gg \ \ll n_g, n_h .. ; n_{\alpha}, n_{\beta} \ldots; \mu
\parallel \eqno(A.5)
$$
$$
 =  \sum_{{\stackrel{n_g,n_h \ldots}{n_{\alpha}, n_{\beta}\ldots}}} \,
\sum_{\lambda, \nu} \vert n_g, n_h \ldots ; n_{\alpha}, {n_\beta} \ldots ;
\nu \rangle \, (M^{-1})_{\nu \lambda} \, \langle n_g \ldots ; n_{\alpha}
\ldots ; \lambda \vert     \eqno(A.6)
$$

\vspace{0.2cm}
\noindent {\underline {Projection Operators :}}
$$
P (n_g,n_h \ldots; n_{\alpha}, n_{\beta} \ldots) \, 
$$
$$
= \, \sum_{\mu} \parallel n_g, n_h \ldots; n_{\alpha}, 
n_{\beta} \ldots ; \mu \gg \, \ll n_g, n_h \ldots ; n_{\alpha}, 
n_{\beta} \ldots; \mu \parallel \eqno(A.7)
$$
$$
= \sum_{\lambda, \nu} \vert n_g, n_h \ldots ; n_{\alpha}, n_{\beta} \ldots ;
\nu \rangle \, (M^{-1})_{\nu \lambda} \, \langle n_g \ldots ; n_{\alpha}
\ldots ; \lambda \vert    \eqno(A.8)
$$
$$
I \, = \, \sum_{{\stackrel{n_g,n_h \ldots}{n_{\alpha}, n_{\beta}\ldots}}}
\, P(n_g, n_h \ldots; n_{\alpha}, n_{\beta} \ldots)    \eqno(A.9) 
$$
$$
P(n_g, n_h \ldots; n_{\alpha}, n_{\beta} \ldots) \, 
\parallel n^{'}_g, n^{'}_h \ldots; n^{'}_{\alpha}, n^{'}_{\beta} \ldots ;
\mu \gg \, 
$$
$$
\quad = \quad \delta_{n_g n^{'}_g} \ldots \delta_{n_{\alpha}
n^{'}_{\alpha}} \ldots \parallel n_g \ldots ; n_\alpha \ldots ; 
\mu \gg \eqno(A.10)
$$
$$
P (n_g \ldots ; n_{\alpha} \ldots ) \, \vert n^{'}_g \ldots ;
n^{'}_{\alpha} \ldots ; \mu \rangle \, 
$$
$$
\qquad = \quad \delta_{n_g n^{'}_g} \ldots
\delta_{n_{\alpha}n^{'}_{\alpha}} \ldots \vert n_g \ldots ; n_{\alpha}
\ldots ; \mu \rangle     \eqno(A.11)
$$

\noindent {\underline {Number operators}}
$$
N_k \, = \, \sum_{{\stackrel{n_g \ldots n_k \ldots}{n_{\alpha} \ldots}}}
\, n_k \, P (n_g \ldots n_k \ldots ; n_{\alpha} \ldots)   \eqno(A.12)
$$
$$
N_{\beta}  \, = \, \sum_{{\stackrel{n_g \ldots}{n_{\alpha}, n_{\beta} \ldots}}}
\, n_{\beta}  \, P (n_g \ldots ; n_{\alpha}, n_{\beta}  \ldots)
\eqno(A.13)
$$
$$
N_k \parallel n_g \ldots n_k \ldots ; n_{\alpha}, n_{\beta} \ldots ; \mu
\gg  
$$
$$
\qquad =  \quad n_k \parallel n_g \ldots n_k \ldots ; 
n_{\alpha}, n_{\beta} \ldots ; \mu \gg \eqno(A.14)
$$
$$
N_k \vert n_g \ldots n_k \ldots ; n_{\alpha}, n_{\beta} \ldots ; \mu
>   =   
n_k \vert  n_g \ldots n_k \ldots ; n_{\alpha}, n_{\beta} \ldots ; \mu
\rangle \eqno(A.15)
$$
$$
N_{\beta} \parallel n_g \ldots n_k \ldots ; n_{\alpha}, n_{\beta} \ldots ; \mu
\gg 
$$
$$
\qquad   =   \quad n_{\beta} \parallel n_g \ldots n_k \ldots ; 
n_{\alpha}, n_{\beta} \ldots ; \mu \gg \eqno(A.16)
$$
$$
N_{\beta} \vert n_g \ldots n_k \ldots ; n_{\alpha}, n_{\beta} \ldots ; \mu
\rangle   =   
n_{\beta} \vert  n_g \ldots n_k \ldots ; n_{\alpha}, n_{\beta} \ldots ; \mu
\rangle  \eqno(A.17)
$$
$$
[N_k, N_j] \, = \, [N_{\alpha}, N_{\beta}] \, = \, [N_k, N_{\alpha}] \,
$$
$$
\qquad 	= \quad	 0 \quad {\rm for \, any } \ k, j 
\ {\rm and \, any} \ \alpha, \beta   \eqno(A.18) 
$$

\noindent Total number operator is
$$
N = \sum_k N_k = \sum_\alpha N_\alpha \eqno(A.19)
$$

\noindent {\underline {Creation and  destruction operators}}  :
$$
c^{\dagger}_{j \beta} \, = \, \sum_{{\stackrel{n_g,n_j
\ldots}{n_{\alpha}, n_{\beta} \ldots}}} \, \sum_{\mu^{'} \nu} \,
A_{\mu^{'} \nu} \, \vert n_g \ldots (n_{j}+1) \ldots ; n_{\alpha},
n_{\beta} + 1 \ldots ; \mu^{'} \rangle \, 
$$
$$
\qquad \otimes \langle n_g \ldots n_j \ldots ; n_{\alpha},n_{\beta} 
\ldots \nu \vert    \eqno(A.20)
$$
$$
{[c^{\dagger}_{j \beta}, N_k]}   =   - c^{\dagger}_{j \beta} 
\delta_{jk} \eqno(A.21)
$$
$$
{[c^{\dagger}_{j \beta}, N_{\alpha}]}   =  - c^{\dagger}_{j \beta} 
\delta_{\alpha \beta}  \eqno(A.22)
$$

\noindent For some particular $\mu$,
$$
\vert 3_g, 2_h; 4_{\alpha}, 1_{\beta} ; \mu \rangle
= \vert 1_{g \alpha} 1_{g \alpha} 1_{g \beta} 1_{h \alpha} 1_{h \alpha}
\rangle \, = \, (c^{\dagger}_{g\alpha})^2 \, c^{\dagger}_{g \beta} \,
(c^{\dagger}_{h\alpha})^2 \, \vert 0 \rangle   \eqno(A.23)
$$
$$
c^{\dagger}_{j\beta} \, \vert n_g \ldots \, n_{j \ldots } ; n_{\alpha},
n_{\beta} \ldots ; \lambda \rangle \, = \, \vert \underbrace{1_{j \beta,}
; n_g \cdots n_j \cdots ; n_{\alpha}, n_{\beta} \cdots ;}_ 
{n_g \cdots n_j +1 \cdots ; n_{\alpha}, n_{\beta} + 1} \lambda \rangle 
\eqno(A.24)
$$
$$
c^{\dagger}_{j \sigma} \, \vert n_g \ldots n_j \ldots ; n_{\alpha}
\ldots n_{\sigma} \ldots ; \lambda \rangle
$$
$$
= \sum_{{\stackrel{n^{'}_g \ldots n^{'}_j \ldots}{n^{'}_{\alpha} \ldots
n^{'}_{\sigma}\ldots}}} \, \sum_{\mu^{'} \nu} \, A_{\mu^{'} \nu} \, \vert
n^{'}_g \ldots (n^{'}_{j}+1) \ldots ; n^{'}_{\alpha} \ldots n^{'}_{\sigma}
+1 \ldots ; \mu^{'} \rangle \eqno(A.25)
$$
$$
\otimes \langle n^{'}_g \ldots n^{'}_j \ldots ; n^{'}_{\alpha} \ldots
n^{'}_{\sigma} \ldots ; \nu \vert  n_g \ldots n_j \ldots
; n_{\alpha} \ldots n_{\sigma} \ldots ; \lambda \rangle
$$
$$
= \sum_{\mu^{'} \nu} \, A_{\mu^{'} \nu} M_{\nu \lambda} \, \vert n_g
\ldots (n_{j}+1) \ldots ; n_{\alpha} \ldots n_{\sigma} +1 \ldots ; \mu^{'}
\rangle \eqno(A.26)
$$
$$
\sum_{\nu} \, A_{\mu^{'}\nu} M_{\nu \lambda} \, = \, \delta_{\mu^{'}
\lambda} \eqno(A.27)
$$
$$
A \, = \, M^{-1} \eqno(A.28)
$$
$$
c^{\dagger}_{j \sigma} \, = \, \sum_{{\stackrel{n_g \ldots
n_j..}{n_{\alpha} .. n_{\sigma}..}}} \, \sum_{\lambda \nu} \,
(M^{-1})_{\lambda \nu} \vert 1_{j\sigma} ; n_g \ldots n_j \ldots ;
n_{\alpha} \ldots n_{\sigma} \ldots ; \lambda \rangle \, 
$$
$$
\qquad \otimes \langle n_g \ldots n_j \ldots ; n_{\alpha} \ldots 
n_{\sigma} \ldots ; \nu \vert \eqno(A.29)
$$

\newpage

\baselineskip=18pt

\noindent {\bf References}

\bigskip

\begin{enumerate}
\item{} L.C.Biedenharn, J. Phys. A {\bf 22} (1989) L873.
\item{} A.J.Mcfarlane, J. Phys. A {\bf 22} (1989) 4581.
\item{} K.Odaka, T.Kishi and S.Kamefuchi, J. Phys. A {\bf
24} (1991) L591.
\item{} R.N.Mohapatra, Physics letters {\bf B 242} (1990) 407.
\item{} M.Arik and D.D.Coon, J. Math. Phys. {\bf 17} (1976) 705.
\item{} S.Chaturvedi and V.Srinivasan, Phys. Rev. {\bf A44} (1992)
8024.
\item{} S.Chaturvedi, A.K.Kapoor, R.Sandhya, V.Srinivasan and R.Simon,
Phys. Rev. {\bf A 43}, (1991) 4555.
\item{} R.Parthasarathy and K.S.Viswanathan, J. Phys. {\bf A 24} (1991)
613.
\item{} K.S.Viswanathan, R.Parthasarathy and R.Jagannathan, J. Phys. A
{\bf 25} (1992) L335.
\item{} M.Arik, Z. Phys. C: Particles and Fields {\bf 51} (1991) 627.
\item{} T.Brzezinski, I.L.Egusquiza and A.J.Macfarlane, Phys. Lett. {\bf
B 311} (1993) 202.
\item{} R.Chakrabarti and R.Jagannathan, J. Phys. {\bf A 25} (1992)
6393.
\item{} O.W.Greenberg, Phys. Rev. Lett. {\bf 64} (1990) 407.
\item{} O.W.Greenberg, Phys. Rev. {\bf D 43} (1991) 4111.
\item{} D.I.Fivel, Phys. Rev. Lett. {\bf 65} (1990) 3361.
\item{} D.Zagier, Commun. Math. Phys. {\bf 147} (1992) 210.
\item{} S.Stanciu, Commun. Math. Phys. {\bf 147} (1992) 147.
\item{} A.K.Mishra and G.Rajasekaran, Pramana - J. Phys. {\bf 38} (1992)
L411.
\item{} A.K.Mishra and G.Rajasekaran, Pramana - J. Phys. {\bf 40} (1993)
149 ; Phys. Lett. A {\bf 188} (1994) 210.
\item{} R.F.Werner, Phys. Rev. {\bf D 48} (1993) 2929.
\item S.L.Woronowicz, Commun. Math. Phys. {\bf 111} (1987) 613 ; Publ.
RIMS Kyoto University {\bf 23} (1987) 117.
\item{} W.Pusz and S.L.Woronowicz, Rep. Math. Phys. {\bf 27} (1989) 231.
\item{} W.Pusz, Rep. Math. Phys. {\bf 27} (1989) 349.
\item M.Chaichian, P.Kulish and J.Lukierski, Phys. Lett. {\bf B262}
(1991) 43.
\item M.Chaichian and P.Kulish, in Nonperturbative methods in
low-dimensional quantum field theories (Proc. 14th John Hopkins Workshop
on Current Problems in Particle Theory, Debrecen, Hungary, 1990) (World
Sci. Pub. Co., River Edge, NJ, 1991) p.213.
\item{} R.Jagannathan, R.Sridhar, R.Vasudevan, S.Chaturvedi,
M.Krishnakumari, P.Santha and V.Srinivasan, J. Phys. A {\bf
25} (1992) 6429.
\item E.G.Floratos, J. Phys. {\bf A 24} (1991) 4739.
\item{} S.K.Soni, J. Phys. A  {\bf 24} (1991) L 169.
\item{} A.Schirrmacher, Zeit. f\"ur Phys.  C: Particles and Fields 
{\bf 50} (1991) 321.
\item B.Zumino, preprint (1991) LBL-30120, UCB-PTH-91/1.
\item A.Shirrmacher, J.Wess and B.Zumino, Zeit. f\'{u}r Phys. {\bf C
49} (1991) 317.
\item L.Hlavaty, Czek. J. Phys. {\bf 42} (1992) 1331.
\item{} D.B.Fairlie and C.K.Zachos, Phys. Lett. {\bf B 256} (1991) 43.
\item{} A.K.Mishra and G.Rajasekaran Preprint IMSc/20-1993.
\item{} H.S.Green, Phys. Rev. {\bf 90} (1953) 270.
\item{} O.W.Greenberg and A.M.L.Messiah, Phys. Rev. {\bf B 136} (1964)
248.     
\item{} Y.Ohnuki and S.Kamefuchi, Quantum Field Theory and
Parastatistics, Springer-Verlag, Berlin-Heidelberg-New York 1982.
\item{} J.Cuntz, Commun. Math. Phys. {\bf 51} (1977) 173.
\item J.M.Leinass and J.Myrheim, Nuov. Cim. {\bf 37B} (1977) 1.
\item F.Wilczek, Phys. Rev. Lett. {\bf 48} (1982) 1144 ; ibid {\bf 49}
(1982) 957.
\item{} G.S.Agarwal and S.Chaturvedi, Mod. Phys. Lett. {\bf A 7} (1992)
2407.
\item C.R.Lee and J.P.Yu, Phys. Lett. {\bf A 150} (1990) 63.
\item{} J.Hubbard, Proc. Roy. Soc. {\bf A 276} (1963) 238.
\item{} J.Hubbard, Proc. Roy. Soc. {\bf A 285} (1965) 542.
\item{} M.Ogata and H.Shiba, Phys. Rev. {\bf B 41} (1990) 2326.
\item{} A.K.Mishra and G.Rajasekaran, Pramana - J. Phys. {\bf 36} (1991)
537 ; ibid {\bf 37} (1991) 455 (E).
\item{} A.K.Mishra and G.Rajasekaran, Mod. Phys. Lett. {\bf A 7} (1992) 3525.
\item{} J.W.Goodison and D.J.Toms, Phys. Rev. Lett. {\bf 71} (1993)
3240.
\item{} Non-perturbative quantum field theory: mathematical aspects and
applications (Selected papers of J\"{u}rg Fr\"{o}hlich), Advanced Series
in Mathematical Physics {\bf 15}, World Scientific 1992.
\item{} K.Fredenhagen, M.Gaberdiel, S.M.R\"{u}ger , Scattering states of
plektons (particles with braid group statistics) in 2+1 dim. QFT,
preprint Hamburg University and DAMTP Cambridge 1994.
\item{} S.V.Shabanov, Preprint BuTP-92/24.
\item S.Iida, Phys. Rev. Lett. {\bf 69} (1992) 1833.
\item T.Suzuki, J. Math. Phys. {\bf 34} (1993) 3453.
\item O.W.Greenberg, Physica {\bf A 180} (1992) 419.
\item{} R.Haag, Local Quantum Physics, Springer 1992.
\end{enumerate}

\newpage


\textheight=20cm
\textwidth=15cm 
\setlength{\oddsidemargin}{-0.75in}
\vfill
\baselineskip=12pt
\begin{center}
\begin{tabular}{|c|l|l|} \hline 
& & \\
\multicolumn{1}{|c|}{\bf Statistics} & \multicolumn{1}{|c|}{\bf Representation} 
&  \multicolumn{1}{|c|}{\bf Algebra}  \\  
& & \\ \hline 
& & \\
Infinite & Standard representation & $c_i c^{\dagger}_j = \delta_{ij}$ \\
&  & \\
'' & q-mutator (with real q) & $c_i c^{\dagger}_j 
-qc^{\dagger}_j c_i  = \delta_{ij}$ \\
& & \\
'' & Two-parameter algebra &  \{ \begin{tabular}{l} 
$c_i c^{\dagger}_j - q_1 c^{\dagger}_jc_i$ \\  $-q_2 \delta_{ij} 
\Sigma_k c^{\dagger}_k c_k = \delta_{ij}$ \end{tabular}  \\
& & \\
'' & q-mutator, transformed & $c_i c^{\dagger}_j - 
c^{\dagger}_j c_i = \delta_{ij} p^{2 \Sigma_{k<i}{N_k}} p^{N_i}$ \\
& & \\ 
'' & $q$-mutator, transformed & \{ \begin{tabular}{l} 
$c_ic^{\dagger}_j - p^{-1} c^{\dagger}_j c_i = 0, \quad {\rm for} \quad 
i \neq j$ \\
$ c_i c^{\dagger}_i - c^{\dagger}_i c_i = p^{N_i}$  
\end{tabular}  \\
& & \\
'' & q-mutator (with complex q) & \{ \begin{tabular}{l} 
$c_ic^{\dagger}_j-qc^{\dagger}_jc_i = 0, \quad {\rm for} \quad i < j$ \\
$ c_i c^{\dagger}_i -pc^{\dagger}_i c_i = 1$ \end{tabular}  \\ 
& & \\
\begin{tabular}{c}Infinite \\ with \\ Pauli principle \end{tabular} 
& Standard representation & \{ \begin{tabular}{l} $c_ic^\dagger_j = 0, 
\quad {\rm for} \quad i \neq j$ \\
$ c_i c^{\dagger}_i + c^{\dagger}_i c_i = 1$  \end{tabular}  \\
& & \\
''
& q-mutator (with real $q$) & \{ \begin{tabular}{l} 
$c_ic^{\dagger}_j-qc^{\dagger}_jc_i = 0, \quad {\rm for} \quad i \ne j$ \\
$ c_i c^{\dagger}_i + c^{\dagger}_i c_i = 1$. \end{tabular}  \\ 
& & \\
''
& q-mutator (with complex $q$) & \{ \begin{tabular}{l} 
$c_ic^{\dagger}_j-qc^{\dagger}_jc_i = 0, \quad {\rm for} \quad i < j$ \\
$ c_i c^{\dagger}_i + c^{\dagger}_i c_i = 1$ \end{tabular}  \\ 
& & \\ \hline
\end{tabular}

\vspace{.8cm}

{\bf Table I. Representations of infinite statistics.}
\end{center}

\newpage


\textheight=20cm
\textwidth=20cm   
\setlength{\oddsidemargin}{-0.75in}
\vfill
\baselineskip=12pt
\begin{center}
{\small {
\begin{tabular}{|c|c|c|l|l|} \hline 
& & & & \\
{\bf Statistics} & {\bf $cc$ algebra} & {\bf $c_i
c_j^\dagger$ algebra} & \multicolumn{1}{|c|}{\bf $c_j c_j^\dagger$ algebra} & 
\multicolumn{1}{|c|}{\bf Remarks} \\
& & {\bf for $i\ne j$} & & \\ 
& & & &  \\ \hline 
& & & & \\
$q$-statistics & $c_i c_j = q^\ast c_j c_i$ & $c_i c^\dagger_j
= q c^\dagger_j c_i$ & $c_j c^\dagger_j - p c^\dagger_j c_j = |q|^{2
\sum_{i<j} N_i} f(N_j)$ & General \\
& for $i < j$ & for $ i < j$ & & representation \\
& & & & \\
'' & $c_i c_j = q^\ast c_j c_i$ & $c_i c^\dagger_j
= q c^\dagger_j c_i$ & 
$c_j c^\dagger_j - p c^\dagger_j c_j = |q|^{2 \sum_{i<j} N_i}
\left\{ 1-p (1-\delta_{N_j,0})\right\}$ & \\
& for $i < j$ & for $ i < j$ & & \\
& & & & \\
'' & $c_i c_j = q^\ast c_j c_i$ & $c_i c^\dagger_j
= q c^\dagger_j c_i$ & 
$c_j c^\dagger_j = |q|^{2 \sum_{i<j} N_i}$ &  \\
& for $i < j$ & for $ i < j$ & & \\
& & & & \\
'' & $c_i c_j = q^\ast c_j c_i$ & $c_i c^\dagger_j
= q c^\dagger_j c_i$ & $c_j c^\dagger_j- pc^\dagger_j c_j = 
|q|^{2 \sum_{i<j} N_i}$ &  \\
& for $i < j$ & for $ i < j$ & & \\
& & & & \\
'' & $c_i c_j = q^\ast c_j c_i$ & $c_i c^\dagger_j
= q c^\dagger_j c_i$ & $c_j c^\dagger_j- |q|^2 c^\dagger_j c_j = 1 +
(|q|^2-1) \sum_{k<j} c^\dagger_k c_k$ & Covariant \\
& for $i < j$ & for $ i < j$ & & under $SU_q(n)$ \\
& & & & \\
bose statistics & $c_i c_j = c_j c_i$ & $c_i c^\dagger_j
= c^\dagger_j c_i$ & $c_j c^\dagger_j- pc^\dagger_j c_j = f(N_j)$
& General \\
& & for $ i \ne j$ & & representation \\
& & & & \\
'' & $c_i c_j = c_j c_i$ & $c_i c^\dagger_j
= c^\dagger_j c_i$ & 
$c_j c^\dagger_j - p c^\dagger_j c_j = \left\{ 1-p (1-\delta_{N_j,0})\right\}$ 
& \\
& & for $ i \ne j$ & & \\
& & & & \\
'' & $c_i c_j = c_j c_i$ & $c_i c^\dagger_j
= c^\dagger_j c_i$ & $c_j c^\dagger_j = 1$ & simplest \\
& & for $ i \ne j$ &  & representation \\
& & & & \\ \hline
\end{tabular}}}

\vspace{1cm}

{\bf Table II. \ Statistics and algebras in the bosonic Fock space.}

\end{center}

\newpage

\noindent {\bf Table II (Continued)}

\vspace{.5cm}

\begin{center}
{\small {
\begin{tabular}{|c|c|c|l|l|} \hline
& & & & \\
bose statistics & $c_i c_j = c_j c_i$ & $c_i c^\dagger_j
= c^\dagger_j c_i$ & $c_j c^\dagger_j - pc^\dagger_j c_j = 1 ; p \ne -1 $
& Commuting \\
& & for $ i \ne j$ & & deformed oscillators \\
& & & & \\
'' & $c_i c_j = c_j c_i$ & $c_i c^\dagger_j
= c^\dagger_j c_i$ & $c_j c^\dagger_j - c^\dagger_j c_j = 1$  
& Canonical rep. \\
& & for $ i \ne j$ & & of bosons \\
& & & & \\
fermi statistics & $c_i c_j = -c_j c_i$ & $c_i c^\dagger_j
= -c^\dagger_j c_i$ & $c_j c^\dagger_j- pc^\dagger_j c_j = f(N_j)$
& General \\
& for $i \ne j$  & for $ i \ne j$ & & representation \\
& & & & \\
'' & $c_i c_j = -c_j c_i$ & $c_i c^\dagger_j
= -c^\dagger_j c_i$ & $c_j c^\dagger_j - p c^\dagger_j c_j = 
\left\{ 1-p (1-\delta_{N_j,0})\right\}$ & \\
& for $i \ne j$  & for $ i \ne j$ & & \\
& & & & \\
'' & $c_i c_j = -c_j c_i$ & $c_i c^\dagger_j
= -c^\dagger_j c_i$ & $c_j c^\dagger_j = 1$ 
& \\
& for $i \ne j$  & for $ i \ne j$ & & \\
& & & & \\
'' & $c_i c_j = -c_j c_i$ & $c_i c^\dagger_j
= -c^\dagger_j c_i$ & $c_j c^\dagger_j - pc^\dagger_j c_j = 1 ; p \ne
-1$ &  Anticommuting \\
& for $i \ne j$  & for $ i \ne j$ & & deformed oscillators \\
& & & & \\
`` & $c_i c_j = -c_j c_i$ & $c_i c^\dagger_j
= -c^\dagger_j c_i$ & $c_j c^\dagger_j - c^\dagger_j c_j = 1$ 
&  Anticommuting \\
& for $i \ne j$  & for $ i \ne j$ & & bosons \\ 
& & & & \\
fractional & $c_i c_j = e^{-i\theta} c_j c_i$ & 
$c_i c^\dagger_j = e^{i\theta} c^\dagger_j c_i$ & 
Any one of the same $c_j c^\dagger_j$ & \\
statistics & for $i < j$ & for $ i < j$ & 
relations given in bose or & \\
& & & fermi statistics above & \\ 
& & & & \\ \hline 
\end{tabular}}}
\end{center}

\newpage


\textheight=20cm
\textwidth=20cm   
\setlength{\oddsidemargin}{-0.75in}
\baselineskip=12pt
\vfill

\begin{center}
{\small {
\begin{tabular}{|c|c|c|l|l|} \hline
& & & & \\
{\bf Statistics} & {\bf $cc$ algebra} & {\bf $c_i
c_j^\dagger$ algebra} & \multicolumn{1}{|c|}{\bf $c_j c_j^\dagger$ algebra} & 
\multicolumn{1}{|c|}{\bf Remarks} \\
& & {\bf for $i\ne j$} & & \\ 
& & & & \\ \hline 
& & & & \\
$q$-statistics & $c_i c_j = q^\ast c_j c_i$ & $c_i c^\dagger_j
= q c^\dagger_j c_i$ & 
$c_j c^\dagger_j - p c^\dagger_j c_j = (\delta_{N_j,0} - p
\delta_{N_j,1})$ & General \\
& for $i < j$ & for $ i < j$ & $ \times \{ |r|^2+(|q|^2-1) \sum_{k<j}
c^\dagger_k c_k\}$ & representation \\
& $c_ic_i = 0$ & & & \\
& & & & \\
'' & $c_i c_j = q^\ast c_j c_i$ & $c_i c^\dagger_j
= q c^\dagger_j c_i$ & $c_jc_j^\dagger + c^\dagger_jc_j=|r|^2+(|q|^2-1)
\sum_{k<j} c_k^\dagger c_k$ & \\ 
& for $i < j$ & for $ i < j$ & & \\
& $c_ic_i = 0$ & & & \\
& & & & \\
'' & $c_i c_j = q^\ast c_j c_i$ & $c_i c^\dagger_j
= q c^\dagger_j c_i$ & $c_j c^\dagger_j + c^\dagger_j c_j = 
1 + (|q|^2-1) \sum_{k<j} c^\dagger_k c_k$ & Covariant \\ 
& for $i < j$ & for $ i < j$ & & under $SU_q(n)$ \\
& $c_ic_i = 0$ & & & \\
& & & & \\
bose-statistics & $c_i c_j =  c_j c_i$ & $c_i c^\dagger_j
= c^\dagger_j c_i$ & $c_j c^\dagger_j - p c^\dagger_j c_j = 
(\delta_{N_j,0} - p \delta_{N_j,1})|r|^2$ & \\ 
& $c_i c_i = 0$ & for $ i \ne j$ & & \\
& & & & \\
'' & $c_i c_j =  c_j c_i$ & $c_i c^\dagger_j
= c^\dagger_j c_i$ & $c_j c^\dagger_j + c^\dagger_j c_j = 1$ 
& Commuting  \\ 
& $c_i c_i = 0$ & for $ i \ne j$ & & fermions \\
& & & & \\
fermi-statistics & $c_i c_j =  -c_j c_i$ & $c_i c^\dagger_j
= -c^\dagger_j c_i$ & $c_j c^\dagger_j - p c^\dagger_j c_j = 
(\delta_{N_j,0} - p \delta_{N_j,1})|r|^2$ & \\ 
& for all $i,j$ & for $i \ne j$  & & \\
& & & & \\
'' & $c_i c_j =  -c_j c_i$ & $c_i c^\dagger_j
= -c^\dagger_j c_i$ & $c_j c^\dagger_j + c^\dagger_j c_j = 1$ 
& Canonical rep. \\ 
&  for all $i,j$ & for $i \ne j$  & & of fermions. \\
& & & & \\
fractional & $c_i c_j = e^{-i\theta} c_j c_i$ & 
$c_i c^\dagger_j = e^{i\theta} c^\dagger_j c_i$ & 
$c_j c^\dagger_j - p c^\dagger_j c_j = 
(\delta_{N_j,0} - p \delta_{N_j,1})|r|^2$ & \\ 
statistics & for $i < j$ & for $ i < j$ & & \\
& $c_i c_i = 0$  & & & \\
& & & & \\
'' & $c_i c_j = e^{-i\theta} c_j c_i$ & 
$c_i c^\dagger_j = e^{i\theta} c^\dagger_j c_i$ & 
$c_j c^\dagger_j + c^\dagger_j c_j = 1$ & \\
& for $i < j$ & for $ i < j$ & & \\
& $c_i c_i = 0$  & & & \\ 
& & & & \\ \hline 
\end{tabular}}}

\vspace{.5cm}

{\bf Table III. \ Statistics and algebras in the fermionic Fock space.}

\end{center}

\newpage


\textheight=20cm
\textwidth=20cm 
\setlength{\oddsidemargin}{-0.75in}
\baselineskip=12pt
\vfill
{\small {
\begin{center}
\begin{tabular}{|c|c|l|} \hline 
& & \\ 
{\bf $N_{k\alpha}$} & 
\parbox[c]{2cm}{\begin{tabular}{c}
{\bf Statistics in} \\ {\bf latin indices} 
\end{tabular}} \hspace{1cm} 
\begin{tabular}{c}
\parbox[c]{3cm}{{\bf Statistics in} \\ {\bf greek indices} }
\end{tabular} & 
\multicolumn{1}{|c|}{\bf $cc^\dagger$ algebra}  \\ 
& & \\ \hline 
& & \\
Do not exist & 
\parbox[c]{2cm}{~~~~~~~~infinite} \hspace{2cm}
\parbox[c]{3cm}{infinite} & 
$c_{k\alpha} c^\dagger_{m\beta} - q \delta_{\alpha\beta} 
\sum_\gamma c^\dagger_{m\gamma} c_{k\gamma} =
\delta_{km} \delta_{\alpha\beta} $ (real $q$) \\
& & \\
'' & 
\parbox[c]{2cm}{~~~~~~~~''} \hspace{2cm}
\parbox[c]{3cm}{~~~~''} & 
$\left\{ \begin{tabular}{ll}
$c_{k\alpha} c^\dagger_{m\beta} - q
\delta_{\alpha\beta} \sum_\gamma c^\dagger_{m\gamma} c_{k\gamma} = 0$ \,,
 & for \ $k < m$ \\  
\\ $c_{k\alpha} c^\dagger_{k\beta} - p
\delta_{\alpha\beta} \sum_\gamma c^\dagger_{k\gamma} c_{k\gamma} = 
\delta_{\alpha\beta}$ ( & complex $q$ \\  & and real $p$) 
\end{tabular} \right.$ \\
& & \\
'' & \begin{tabular}{c} 
\parbox[c]{2cm}{~~~~~~~Fermi/} \\ 
\parbox[c]{2cm}{~~~~~~~Bose} \end{tabular} \hspace{2cm}
\parbox[c]{3cm}{~~~''} & 
$c_{k\alpha} c^\dagger_{m\beta} \pm 
\delta_{\alpha\beta} \sum_\gamma c^\dagger_{m\gamma} c_{k\gamma} =
\delta_{km} \delta_{\alpha\beta} $  \\
& & \\
'' & \begin{tabular}{c} 
\parbox[c]{2cm}{~~~~~~~$q$-fermi/} \\
\parbox[c]{2cm}{~~~~~~$q$-bose} \end{tabular} \hspace{2cm}  
\parbox[c]{3cm}{~~~''} & 
$\left\{ \begin{tabular}{l}
$c_{i\alpha} c^\dagger_{j\beta} - q
\delta_{\alpha\beta} \sum_\gamma c^\dagger_{j\gamma} c_{i\gamma} = 0
\quad$ for $\quad i < j$ \\ \\
$c_{j\alpha} c^\dagger_{j\beta} - x 
\delta_{\alpha\beta} \sum_\gamma c^\dagger_{j\gamma} c_{j\gamma}$ \\ \\
$\quad \quad = \delta_{\alpha\beta} + \delta_{\alpha\beta} (|q|^2-1)
\sum_{k<j} \sum_\gamma c^\dagger_{k\gamma} c^\dagger_{k\gamma}$ \\ \\
$x = |q|^2$  \ for \ $q$-bose ; $x=-1$ \ for \ $q$-fermi 
\end{tabular} \right.$ \\
& & \\
Exist & \begin{tabular}{c} Antisymmetric for total exchange, \\
but $n_k \le 1$ \end{tabular} &
\begin{tabular}{l}	
$c_{k\alpha} c^\dagger_{m\beta} + (1-\delta_{km}) c^\dagger_{m\beta} 
c_{k\alpha}$ \\ \\ $\quad \quad = \delta_{km} \delta_{\alpha\beta} 
\left( 1 - \sum_\gamma c^\dagger_{k\gamma} c_{k\gamma} \right)$ 
\end{tabular} \\
& & \\
'' & \begin{tabular}{c} Symmetric for total exchange for \\
$k \ne m$, but infinite statistics \\ in greek indices for $k = m$ \\ 
\end{tabular} &
\begin{tabular}{l}	
$c_{k\alpha} c^\dagger_{m\beta} - (1-\delta_{km}) c^\dagger_{m\beta} 
c_{k\alpha}$ \\ \\ $\quad \quad = \delta_{km} \delta_{\alpha\beta} 
\left( 1 + \sum_\gamma c^\dagger_{k\gamma} c_{k\gamma} \right)$ 
\end{tabular} \\
& & \\
'' & \begin{tabular}{c} Symmetric for total exchange but \\ 
$n_k \le 1$ \end{tabular} &
\begin{tabular}{l}	
$c_{k\alpha} c^\dagger_{m\beta} - (1-\delta_{km}) c^\dagger_{m\beta} 
c_{k\alpha}$ \\ \\ $\quad \quad = \delta_{km} \delta_{\alpha\beta} 
\left( 1 - \sum_\gamma c^\dagger_{k\gamma} c_{k\gamma} \right)$ 
\end{tabular} \\ 
& & \\ \hline
\end{tabular} 

\vspace{1cm}

{\bf Table IV. Statistics and algebras for 2-indexed
systems.}
\end{center}

\newpage

\baselineskip=24pt
\pagestyle{empty}

\noindent {\bf Figure Captions }

\begin{description}
\item[{Fig.1 }] Inversion diagram for the permutation (213) $\rightarrow $
(312). \\
Positive inversions : (1,3) $\rightarrow$ (3,1) and (2,3) $\rightarrow$
(3,2). \\
Negative inversion : (2,1) $\rightarrow$ (1,2).
\item[{Fig.2}] The complex $q$-plane of the $q$-mutator algebra. The disc
$|q|<1$ corresponds to infinite statistics and the circle $|q|=1$
corresponds to fractional statistics. $F$ and $B$ are the Fermi-Dirac
and Bose-Einstein points.
\item[{Fig.3}] Generalized Fock spaces, quantum statistics, algebras and
their interconnections.
\item[{Fig.4}] Same as Fig.3 for systems with two indices that cannot be
mapped into a single index.
\end{description}
\newpage
\pagestyle{empty}
\begin{figure}[htb]
\begin{center}
\mbox{\epsfig{file=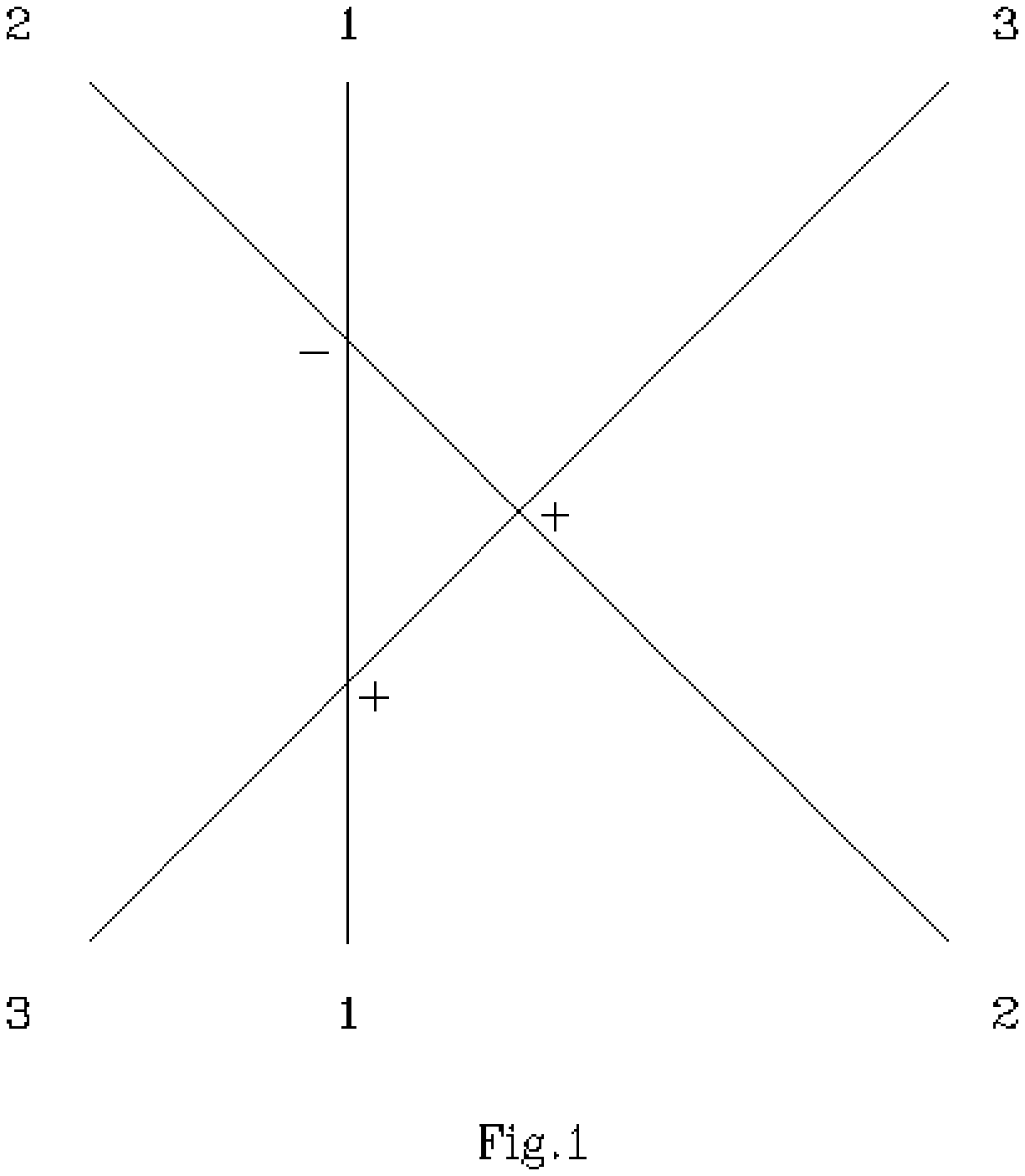,width=10truecm,angle=0}}
\label{Fig1}
\end{center}
\end{figure}
\newpage

\pagestyle{empty}
\begin{figure}[htb]
\begin{center}
\mbox{\epsfig{file=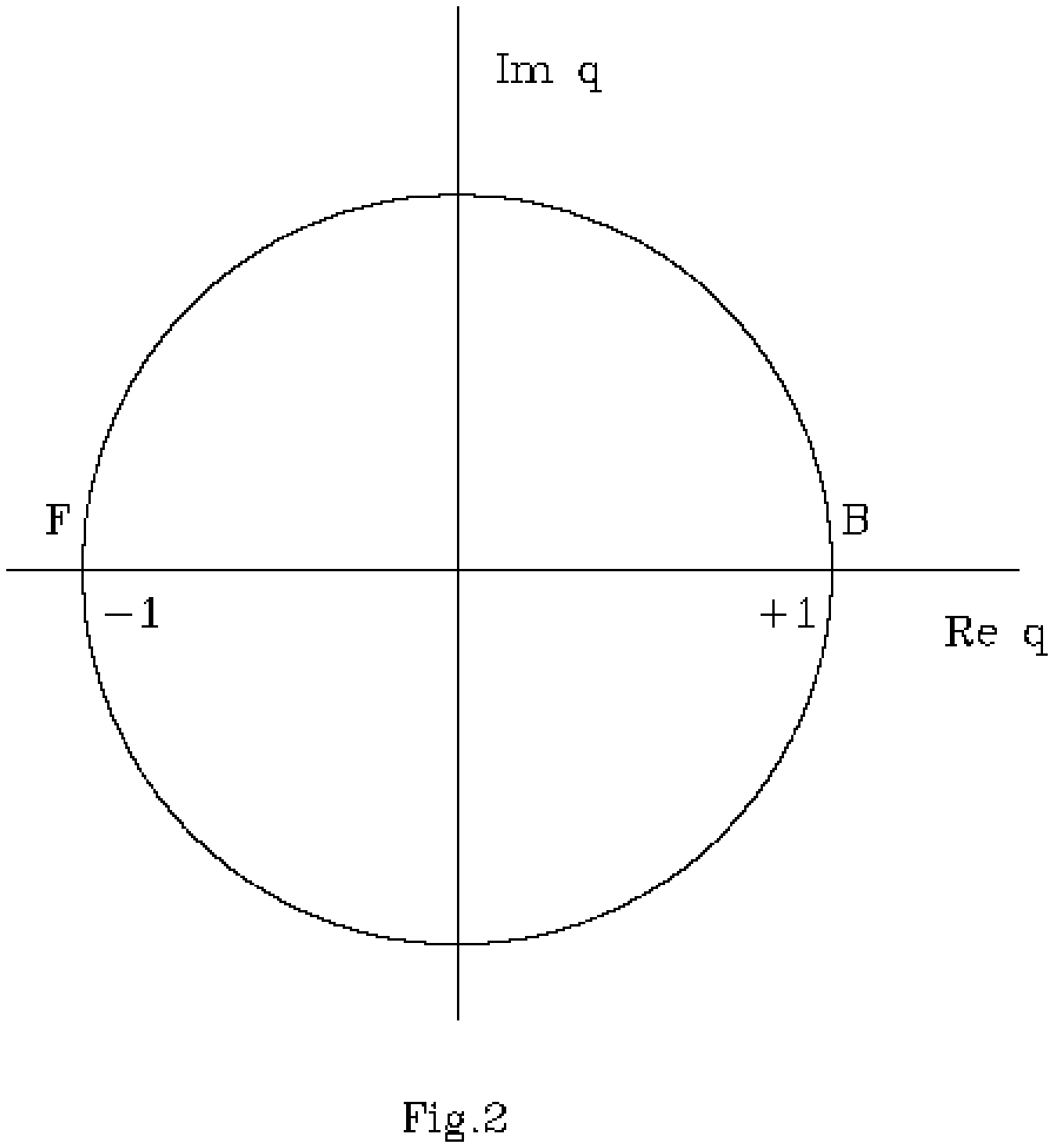,width=10truecm,angle=0}}
\label{Fig2}
\end{center}
\end{figure}
\newpage

\pagestyle{empty}
\oddsidemargin=1cm
\evensidemargin=-.5cm
\thispagestyle{empty}
\begin{picture}(500,450)
\put(60,437.5){\bf GENERALISED     FOCK     SPACES}
\put(-100,430){\line(380,0){380}} 
\put(-100,430){\line(0,20){20}} 
\put(-100,450){\line(380,0){380}} 
\put(280,448){ - - - - - - - - - - } 
\put(280,428){ - - - - - - - - - - } 
\put(365,450){\line(80,0){80}} 
\put(365,430){\line(80,0){80}} 
\put(445,430){\line(0,20){20}} 
\put(-55,430){\vector(0,-60){60}}
\put(102,430){\vector(0,-60){60}}
\put(250,430){\vector(0,-60){60}}
\put(400,430){\vector(0,-60){60}}


\put(-100,295){\framebox(90,75) {\parbox[t]{2.5cm}{Fock space with frozen
order}}}

\put(57,295){\framebox(90,75) {\parbox[t]{2.5cm}{Bosonic and Fermionic
spaces}}}
\put(205,295){\framebox(90,75) {\parbox[t]{2.5cm}{Parafermionic and
Parabosonic spaces}}}
\put(355,295){\framebox(90,75) {\parbox[t]{2.0cm}{Super Fock space}}}
\put(-55,295){\vector(0,-60){60}}

\put(102,295){\vector(0,-30){60}}

\put(250,295){\vector(0,-60){60}}
\put(400,295){\vector(0,-60){60}}
\put(30,265){\line(140,0){140}}
\put(30,265){\vector(0,-30){30}}
\put(170,265){\vector(0,-30){30}}
\put(250,295){\vector(0,-60){60}}
\put(212,265){\line(76,0){76}}
\put(212,265){\vector(0,-30){30}}
\put(250,265){\vector(0,-30){30}}
\put(288,265){\vector(0,-30){30}}
\put(-100,160){\framebox(80,75) {\parbox[t]{1.8cm}{Null Statistics}}}
\put(00,160){\framebox(60,75) {\parbox[t]{2.0cm}{q-statistics}}}
\put(70,160){\framebox(60,75) {\parbox[t]{1.6cm}{Bose or Fermi
statistics}}}
\put(140,160){\framebox(60,75) {\parbox[t]{1.6cm}{Fractional
statistics}}}
\put(220,160){\framebox(60,75) {\parbox[t]{1.6cm}{Para-statistics}}} 
\put(355,160){\framebox(60,75) {\parbox[t]{1.6cm}{  Infinite   statistics}}} 
\put(-55,160){\vector(0,-60){60}}
\put(30,160){\vector(0,-30){30}}
\put(0,130){\line(65,0){65}}
\put(0,130){\vector(0,-30){30}}
\put(65,130){\vector(0,-30){30}}
\put(170,160){\vector(0,-60){60}}
\put(250,160){\vector(0,-60){60}}
\put(211,130){\line(74,0){74}}
\put(211,130){\vector(0,-30){30}}
\put(285,130){\vector(0,-30){30}}
\put(385,160){\vector(0,-60){60}}
\put(320,130){\line(130,0){130}}
\put(320,130){\vector(0,-30){30}}
\put(450,130){\vector(0,-30){30}}
\put(112.5,160){\vector(0,-160){160}}
\put(15,0){\line(210,0){195}}
\put(15,0){\vector(0,-30){30}}
\put(80,0){\vector(0,-30){30}}
\put(145,0){\vector(0,-30){30}}
\put(210,0){\vector(0,-30){30}}
\put(-100,25){\framebox(60,75) {\parbox[t]{1.7cm}{Algebra of Null Statistics}}}
\put(-30,25){\framebox(60,75) {\parbox[t]{1.7cm}{Quantum-group based ~~~~~~~~
algebra}}}
\put(35,25){\framebox(60,75) {\parbox[t]{1.7cm}{Many representations  of
q- statistics}}}
\put(140,25){\framebox(60,75) {\parbox[t]{1.7cm}{Algebras      of fractional
statistics}}}

\put(220,25){\framebox(60,75) {\parbox[t]{1.7cm}{Green's trilinear
algebra}}}

\put(290,25){\framebox(60,75) {\parbox[t]{1.7cm}{Standard
representation}}}
\put(355,25){\framebox(60,75) {\parbox[t]{2.0cm}{q-mutator with real
q}}}
\put(420,25){\framebox(60,75) {\parbox[t]{2.0cm}{q-mutator with complex q}}}
\put(-15,-105){\framebox(60,75) {\parbox[t]{1.7cm}{Canonical bosonic and
fermionic algebras}}}
\put(50,-105){\framebox(60,75) {\parbox[t]{1.7cm}{Deformed
oscillators}}}
\put(115,-105){\framebox(60,75) {\parbox[t]{2.0cm}{Commuting fermions}}}
\put(180,-105){\framebox(60,75) {\parbox[t]{1.7cm}{Anti- commuting
bosons}}}

\put(150,-175){\bf Fig.3}
\end{picture}
\newpage

\pagestyle{empty}
\begin{picture}(500,450)
\put(-100,435){\framebox(540,15){\bf Generalised 
 Fock    Spaces    with    two
   indices}}

\put(-55,435){\vector(0,-70){60}}
\put(102,435){\vector(0,-70){60}}
\put(235,435){\vector(0,-70){60}}
\put(395,435){\vector(0,-70){60}}


\put(-100,295){\framebox(90,80) {\parbox[t]{2cm}{Fock space with new
exclusion principle}}}
\put(57,295){\framebox(90,80) {\parbox[t]{2cm}{Fock space with new
"inclusion" principle}}}

\put(190,295){\framebox(90,80) {\parbox[t]{2cm}{Orthobosonic and orthofermionic
spaces}}}
\put(350,295){\framebox(90,80) {\parbox[t]{2.0cm}{Super Fock spaces
with two decoupled indices}}}

\put(-55,295){\vector(0,-70){37.50}}
\put(102,295){\vector(0,-70){75}}
\put(235,295){\vector(0,-70){37.50}}
\put(395,295){\vector(0,-70){75.50}}
\put(-70,257.5){\line(70,0){70}}
\put(-70,257.5){\vector(0,-70){37.50}}
\put(00,257.5){\vector(0,-37){37.5}}
\put(200,257.5){\line(70,0){70}}
\put(200,257.5){\vector(0,-70){37.50}}
\put(270,257.5){\vector(0,-70){37.50}}
\put(-100,140){\framebox(60,80) {\parbox[t]{1.7cm}{Hubbard statistics}}}
\put(-30,140){\framebox(60,80) {\parbox[t]{1.7cm}{Symmetric Hubbard
statistics}}}
\put( 72,140){\framebox(60,80) {\parbox[t]{1.7cm}{Inclusive
statistics}}}
\put(170,140){\framebox(60,80) {\parbox[t]{1.7cm}{Ortho-
statistics}}}
\put(240,140){\framebox(60,80) {\parbox[t]{1.7cm}{q-ortho-
statistics}}} 
\put(380,140){\framebox(60,80) {\parbox[t]{1.7cm}{Doubly-infinite statistics}}} 
\put(-70,140){\vector(0,-70){75}}
\put(00,140){\vector(0,-37){75}}
\put(102,140){\vector(0,-70){75}}
\put(200,140){\vector(0,-70){75}}
\put(270,140){\vector(0,-70){75}}
\put(395,140){\vector(0,-70){37.5}}
1\put(340,102.5){\line(70,0){70}}
\put(340,102.5){\vector(0,-70){37.5}}
\put(410,102.5){\vector(0,-70){37.5}}
\put(-100,-15){\framebox(60,80) {\parbox[t]{1.7cm}{Hubbard algebra}}}
\put(-30,-15){\framebox(60,80) {\parbox[t]{1.7cm}{Symmetric Hubbard algebra}}}

\put(72,-15){\framebox(60,80) {\parbox[t]{1.7cm}{Inclusive algebra}}}
\put(170,-15){\framebox(60,80) {\parbox[t]{1.7cm}{Standard algebra of
orthostatistics}}}
\put(240,-15){\framebox(60,80) {\parbox[t]{1.7cm}{Generali- sation of
quantum-group algebras}}}
\put(310,-15){\framebox(60,80) {\parbox[t]{1.7cm}{Real-q
representation}}}
\put(380,-15){\framebox(60,80) {\parbox[t]{1.7cm}{Complex-q
representation}}}
\put(150,-125)
{\bf Fig.4}
\end{picture}
\end{document}